\documentclass[journal]{IEEEtran}
\IEEEoverridecommandlockouts

\usepackage{enumitem}
\usepackage{pifont} 
\usepackage{amsmath,amssymb,amsfonts,amsthm}
\usepackage{accents}
\usepackage{algorithmic}
\usepackage{cuted}
\usepackage{breqn}
\usepackage{balance,cite}
\usepackage{graphicx}
\usepackage{tikz}
\usepackage{xcolor}
\usepackage[hidelinks]{hyperref}
\usepackage{subcaption}
\usepackage[font=footnotesize]{caption}

\newtheorem{remark}{\it  Remark}

\AtBeginDocument{}

\newcommand{\mat}[1]{{\mathbf{#1}}}

\newcommand{\matsizec}[2]{\in\mathbb{C}^{#1\times #2}}
\newcommand{\matsizer}[2]{\in\mathbb{R}^{#1\times #2}}

\newcommand{\abs}[1]{\big|#1\big|}

\newcommand{\meter}{$\text{m}$}
\newcommand{\rdm}[1]{\mathbf{\Sigma}(#1)}
\newcommand{\doubleFourier}[3]{\mat{F}_{#1}#2\mat{F}_{#3}^{\mathsf{H}}}
\newcommand{\fmat}[1]{\mat{F}_{#1}}
\newcommand{\hdiv}{\oslash}

\newcommand{\peak}[1]{$\langle #1 \rangle$}

\newcommand{\review}[1]{\textcolor{black}{#1}}

\title{OFDM-based JCAS under Attack: The Dual Threat of Spoofing and Jamming in WLAN Sensing}
\author{Hasan Can Yildirim,~\IEEEmembership{Member,~IEEE},  
Musa Furkan Keskin,~\IEEEmembership{Member,~IEEE},\\ 
Henk Wymeersch,~\IEEEmembership{Fellow,~IEEE}, 
Fran{\c{c}}ois Horlin,~\IEEEmembership{Member,~IEEE}
\thanks{Hasan Can Yildirim (hasan.can.yildirim@ulb.be) and Fran{\c{c}}ois Horlin are with the Wireless Communications Group, Universit{\'e} Libre de Bruxelles, Belgium. Musa Furkan Keskin and Henk Wymeersch are with the Department of Electrical Engineering, Chalmers University of Technology, Sweden.}
\thanks{This work was supported, in part, by the European Commission through the Horizon Europe/JU SNS project Hexa-X-II (Grant Agreement no. 101095759), in part by the Swedish Research Council (VR grant 2023-03821), and part by the Chalmers Transport Area of Advance project \textit{Towards a Multi-Layer Security Vision for Transportation Systems in the 6G Era}.}}

\begin{document}
\maketitle

\begin{abstract}
This study reveals the vulnerabilities of Wireless Local Area Networks (WLAN) sensing, under the scope of joint communication and sensing (JCAS), focusing on target spoofing and deceptive jamming techniques. We use orthogonal frequency-division multiplexing (OFDM) to explore how adversaries can exploit WLAN's sensing capabilities to inject false targets and disrupt normal operations. Unlike traditional methods that require sophisticated digital radio-frequency memory hardware, we demonstrate that much simpler software-defined radios can effectively serve as deceptive jammers in WLAN settings. Through comprehensive modeling and practical experiments, we show how deceptive jammers can manipulate the range-Doppler map (RDM) by altering signal integrity, thereby posing significant security threats to OFDM-based JCAS systems. Our findings comprehensively evaluate jammer impact on RDMs and propose several jamming strategies that vary in complexity and detectability. 
\end{abstract}
\section{Introduction}
\begin{figure}
    \centering
    \includegraphics[width=\columnwidth]{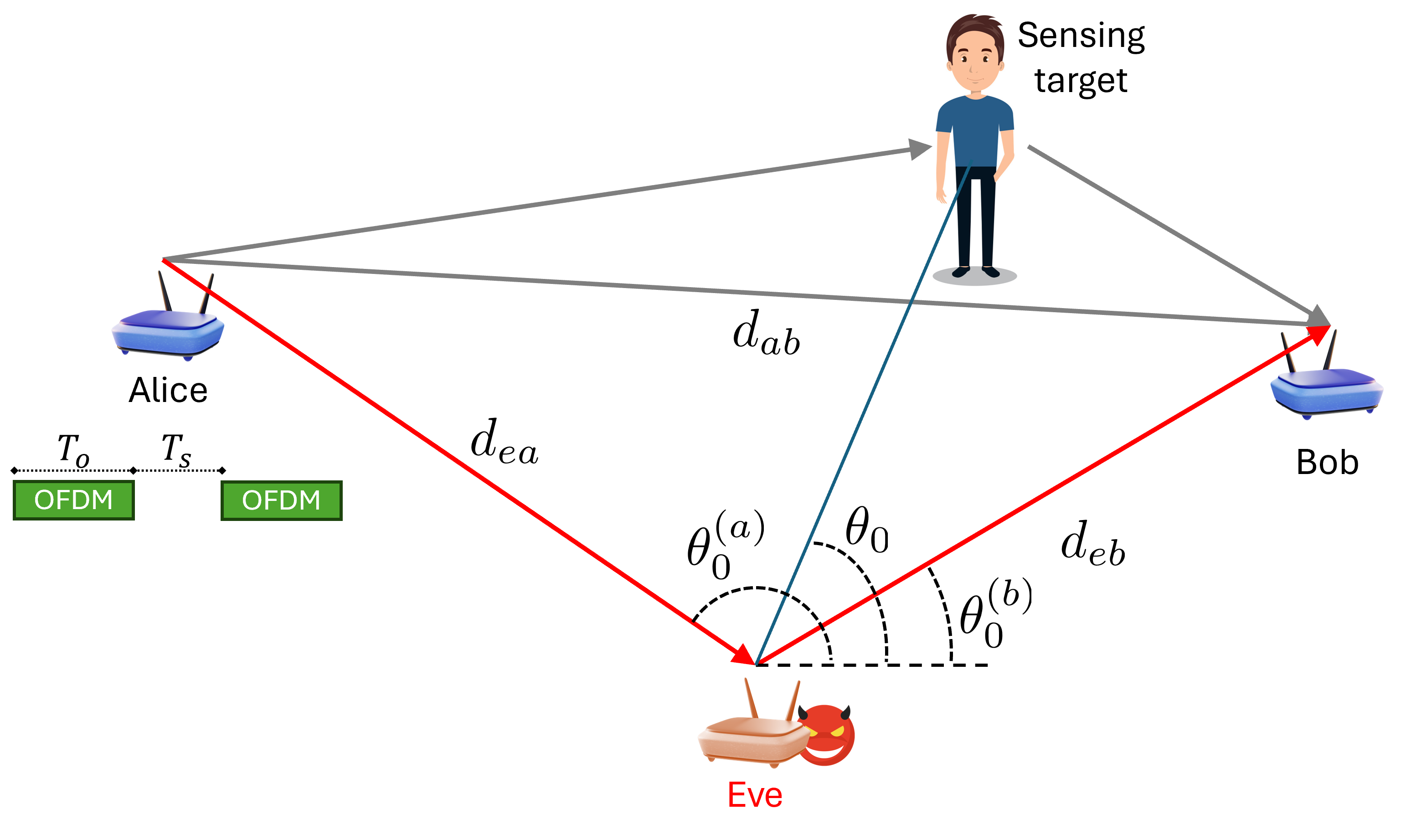}
    \caption{Jammer scenario topology with relevant line-of-sight (LOS) distances $d_{ab}$, $d_{ea}$, and $d_{eb}$, and LOS angles $\theta_0^{(a)}$, $\theta_0^{b}$, and $\theta_0$. Alice operates as a pulsed radar by continuously transmitting the same OFDM symbol of duration $T_o$, with a PRI of $T_s\gg T_o$.}
    \label{fig:scenario_topology}
\end{figure}

Wireless local area network (WLAN) sensing \cite{du21} is a pioneering technology within joint communication and sensing (JCAS) \cite{zhang21a}. It enables WLAN devices to detect, track, and interpret their environment through radio signals. By leveraging orthogonal frequency-division multiplexing (OFDM), WLAN sensing provides channel measurements that are highly applicable to use cases like indoor localization \cite{wlan_sensing_scenarios}, where two devices, Alice and Bob, alternately function as the transmitter and receiver in a half-duplex mode. In this configuration, as illustrated in \autoref{fig:scenario_topology}, the line-of-sight (LOS) between Alice and Bob serves as a critical timing reference, while echoes from surrounding targets, called surveillance signals, are received and processed by Bob for sensing purposes.

\review{However, WLAN sensing, initially designed as a communication-centric technology, was not built with sensing as its primary focus. Sensing features were integrated later, leading to certain vulnerabilities in the system due to its communication-centric design approach. These vulnerabilities present potential entry points for attackers, making WLAN sensing systems increasingly susceptible to security threats. Attackers could exploit these weaknesses to compromise the system, leading to false readings, data manipulation, or disruption of the sensing function altogether. An attacker, Eve, with sophisticated capabilities could eavesdrop on the sensing-related information and transmit jamming signals, severely distorting the received signals at Bob. These attacks are divided into two categories based on their outcome.}

\review{\emph{Target spoofing}, also known as the preamble or fake-path injection \cite{zhang21b}, is the first category where artificial targets are injected at Bob. The literature is focused on exploiting the vulnerabilities of Wi-Fi frame detection, synchronization, and channel estimation \emph{to reduce the communication throughput}. In \cite{zhang21b}, the authors show that joint time and frequency synchronization (JTFS) makes OFDM-based systems vulnerable to attacks. In \cite{lapan13, lapan12}, the frequency and time acquisition algorithms are independently shown to be vulnerable to attacks, and the receiver can be deceived by pilot injection. In \cite{clancy11, patwardhan14}, vulnerabilities in the channel state information feedback mechanism are exploited. In \cite{zhao19}, the orthogonality of OFDM subcarriers is sabotaged by forcing frequency shifts to the subcarriers. Authors in \cite{merwe18} focus on various spoofing attacks, their success rates, and their classifications based on deployment architectures. In \cite{matte15}, how Wi-Fi geolocation spoofing can link devices to their identities is demonstrated. Attacks on public WLAN positioning systems have been explored in \cite{tippenhauer09}, showing how precise models can be used to spoof location information. Methods to identify location spoofing devices in wireless networks are investigated in \cite{liu17}, especially when LOS is absent. From a broader view, surveys on OFDM-based network vulnerability can be found in \cite{pirayesh22, gunther14}, as well as a survey on the physical layer security aspects of OFDM signals in \cite{melki19}.}

\review{\emph{Deceptive jamming} is the second category where the perception of real targets is altered. However, the literature on OFDM-based deceptive jamming is quite sparse. Authors in \cite{schuerger08} designed a deceptive jammer against OFDM-based imaging radars which altered the radar image obtained by the receiver. Meanwhile, \cite{schuerger09} has shown that randomly generated OFDM signals can be further exploited in deceptive jamming. Authors in \cite{tan21} designed a novel method to jam frequency diverse arrays under the narrow band assumption. In \cite{sun18, ji24}, authors investigate deceptive jamming methods for both static and dynamic objects under the imaging radar framework. In \cite{yang22}, an algorithm for the fast generation of deceptive signals is designed. Traditionally, deceptive jamming relied on advanced digital radio-frequency memory (DRFM) hardware to estimate system parameters and mimic targets, which is costly and complex for civil use \cite{drfm_main}. However, WLAN sensing's standardized OFDM symbols enable simpler software-defined radios (SDRs) to perform target spoofing and deceptive jamming without DRFM.}

Several countermeasures have been proposed to address these vulnerabilities in OFDM-based JCAS systems. One approach to combating jammers mimicking authorized signals is secure OFDM precoding, as explored in \cite{liang20}. Other studies, such as \cite{fang24}, propose novel designs like time-frequency modulation using metasurfaces to create deceptive targets in radar profiles. Angle-of-arrival (AoA) based physical-layer authentication, developed in \cite{srinivasan24}, can filter out jammers, while machine and deep learning techniques have been applied to detect 5G signal jammers in \cite{varotto24a, varotto23, varotto24b}. These approaches show promise, but no single method has emerged as universally effective against all types of jamming signals under every scenario. On the other hand, authors in \cite{li24a, li24b} have shown that fake path injection can be used to increase location privacy in wireless networks. While authors in \cite{argyriou23} focus on preventing passive emitter tracking in OFDM-based systems, and in \cite{schuerger09} on the effectiveness of randomly generated OFDM waveforms against deceptive jamming.

\review{In this paper, we explore SDRs in WLAN sensing and the security risks they pose on future OFDM-based JCAS systems. While this study focuses on the sensing aspect, it is crucial to point out that this is within the broader context of communication-centric JCAS systems. We believe that our study provides valuable insights into how the reuse of communication systems for sensing opens up new threat vectors, which future JCAS designs must consider, ensuring both secure communication and accurate sensing. We build upon and significantly extend the research originally introduced in our previous conference paper \cite{yildirim24}, by providing a comprehensive analysis of both target spoofing and deceptive jamming strategies, offering greater control and predictability. By incorporating carrier frequency offset (CFO) into our models, we demonstrate that once Bob synchronizes with Eve's signal, Alice's signal is fully invalidated, ensuring a more deterministic impact on the attacking process for deceptive jamming. By combining various methods that exploit Wi-Fi standardization, advanced jamming strategies are introduced for i) artificial target injection such as overcrowding, selective target injection, and advanced target mimicry, and ii) invalidating the surveillance signal with the preceding jamming signal or the forced synchronization. We also provide an in-depth numerical evaluation of their complexity and effectiveness. Furthermore, the experimental validation is enhanced by implementing these more sophisticated jamming strategies.}
While giving the jammer extensive capabilities, we maintain the WLAN sensing framework in its current form and focus on RDM-based sensing. This methodology is intended to reveal the vulnerabilities inherent in WLAN sensing systems, and more broadly, in all OFDM-based JCAS systems, when subjected to target spoofing and deceptive jamming. 

\subsection{Contributions}

Our primary contributions are summarized as follows:
\begin{itemize}
    \item \textbf{Spoofing and Jamming Framework:} We introduce a detailed mathematical framework for target spoofing and deceptive jamming in WLAN sensing systems. We begin with a basic single-antenna, single-target jamming model that highlights the intricacies of signal generation and the interaction between the jamming and legitimate sensing signals. This foundational model is expanded to include more sophisticated scenarios involving multiple antennas and advanced target mimicry techniques. We show that these enhancements allow the jammer to more effectively disrupt the sensing process, significantly increasing the impact of the attack and the jammer's complexity. 
    \item \textbf{Evaluation of Strategies and Their Implications:}
    We systematically evaluate various jamming strategies, considering their complexity, effectiveness, and detectability. Our analysis covers both simple jamming techniques and more advanced methods that exploit vulnerabilities in Wi-Fi standardization. We demonstrate how these strategies can distort RDMs and affect the target detection probabilities, compromising the integrity of the sensing process and posing significant security threats.
    \item \textbf{Experimental Validation with SDR Platforms:}
    We implement the proposed jamming strategies using SDR platforms to validate our theoretical findings. Our experiments confirm the feasibility of executing sophisticated jamming attacks with relatively simple and accessible hardware, highlighting the real-world applicability of our approach and the pressing need for enhanced security measures in WLAN sensing systems.
\end{itemize}

\subsection{Related Work}
\review{In \autoref{tab:literature_comparison}, the comparison between the relevant works and this study is provided. Authors in \cite{zhang21b,lapan13,lapan12,clancy11,patwardhan14,zhao19} explored jamming strategies that target the communication throughput. To do so, they exploited the JTFS, independent frequency/time synchronization, and known signal structures. Since the focus was on communication throughput, the key performance indicator (KPI) for jamming was mainly the bit-error rate (BER). In \cite{schuerger09}, authors targeted an imaging radar by exploiting the channel estimation process, and they showed a reduction in target probability of detection (PD). As mentioned earlier, our previous work \cite{yildirim24} targeted WLAN sensing by exploiting only the time synchronization (TS) and the only KPI was RDM. 
In \cite{yildirim24}, we discovered the fundamental vulnerabilities of WLAN sensing. The target spoofing was achieved by transmitting OFDM symbols modulated with artificial channel transfer functions. Meanwhile, the deceptive jamming performance was dependent on the time alignment between the cyclic prefix of the legitimate and jamming signals. Eve could either range-shift the true RDM observed by Bob or destroy the subcarrier orthogonality of Alice's signals, effectively turning them into noise. However, these effects were uncontrolled by Eve and occurred randomly. 
In this study, we continue to target WLAN sensing by exploiting JTFS, known signal structures, and channel estimation procedures. We show RDMs for illustrative purposes, and the main KPI is the target PD.}

The remainder of this paper is organized as follows. WLAN sensing framework, the scenario, system model, and radar processing are described in \autoref{sec:wlan_sensing}. The basic jammer functionalities, such as signal generation, jammer channel, and the jammed RDM are modeled in \autoref{sec:deceptive_jammer}. Based on these analyses, various advanced jamming strategies and their complexity and effectiveness are discussed in \autoref{sec:jamming_strategies}. Numerical analyses are provided in \autoref{sec:numerical_analyses}, which are experimentally validated in \autoref{sec:experimental_validation}. Finally, the conclusion is drawn in \autoref{sec:conclusion}.

\begin{table}[]
\centering
\begin{tabular}{l|lll}
\textbf{Ref.} & \textbf{Target}     & \textbf{Exploit}   & \textbf{KPI} \\ \hline
\cite{zhang21b}      & Throughput           & JTFS               & BER          \\
\cite{lapan13}      & Throughput           & FS                 & FER          \\
\cite{lapan12}      & Throughput           & TS                 & TER          \\
\cite{clancy11}      & Throughput           & Known signals      & BER          \\
\cite{patwardhan14}      & Throughput           & Known signals      & BER          \\
\cite{zhao19}     & Throughput           & JTFS               & BER          \\ 
\cite{schuerger09}     & Imaging              & Channel estimation & Target PD    \\
\cite{yildirim24}     & Sensing & TS & RDM \\ \hline 
This         & Sensing              & JTFS               & RDM          \\
study        &                      & Known signals      & Target PD    \\
             &                      & Channel estimation &
\end{tabular}
\caption{\label{tab:literature_comparison} The acronyms are FS: frequency synchronization, TS: time synchronization, FER: Frequency synchronization error rate, TER: time synchronization error rate, RDM: range-Doppler map, PD: probability of detection.}
\end{table}

\subsubsection*{Notation} Matrices and vectors are given by bold characters, $\mat{X}$ and $\mat{x}$, respectively. The Hadamard product $\odot$ and division $\hdiv$ correspond to the element-wise multiplication and division between matrices or vectors of the same size, respectively. The forward and inverse Fourier transform matrices of size $N$ are defined as $\mat{F}_N$ and $\mat{F}_N^{\mathsf{H}}$, respectively, where $\mat{F}^{\mathsf{H}}_N$ is the Hermitian transpose of $\mat{F}_N$. Finally, the phase of a cisoid is obtained as $\angle(e^{-jX})=-X$.

\section{System Model}
\label{sec:wlan_sensing}
In this section, we describe the high-level jammer scenario. Then, we detail the sensing framework and signal processing chain under nominal operating conditions (i.e., without a jammer) \review{where we assume that Bob employs RDM-based processing}. Finally, we identify the main vulnerabilities in WLAN sensing.

\subsection{Scenario Description}
\label{sec:scenario_description}
The scenario consists of three stationary devices, Alice, Bob, and Eve, and a mobile target, as shown in \autoref{fig:scenario_topology}. Alice and Bob are involved in the legitimate WLAN Sensing, while Eve acts as the jammer. 
\autoref{tab:scenario_view} details the operating modes of the devices during different stages of the process.

\subsubsection{WLAN Sensing}
Alice and Bob operate in half-duplex mode, alternating as transmitter (Tx) and receiver (Rx) using OFDM for channel measurements in a bistatic setup. They first exchange sensing parameters during the \emph{negotiation} phase and localize each other while exchanging their roles. Then, Alice sends sensing signals during the \emph{measurement} phase, which Bob receives and processes the surveillance signal for sensing. Bob uses RDM-based processing for sensing, the most common approach.

\subsubsection{Jamming}
While WLAN sensing is taking place, an adversarial device, Eve, intervenes with the WLAN sensing procedure. It transmits signals that can introduce artificial targets to Bob, and potentially invalidate the surveillance signal perceived by Bob. To achieve these, Eve has to operate in a dual capacity: i) as an Rx during sensing negotiation between Alice and Bob to \emph{eavesdrop} on the sensing parameters, and to potentially deduce other metrics related to the topology, and ii) as a Tx during a sensing measurement to emit \emph{spoofing and jamming} signals targeting Bob.

\begin{table}[t]
\caption{Each device's role during a given stage.\label{tab:scenario_view}}
\centering
\begin{tabular}{rccc}
\textit{WLAN Sensing}   & \multicolumn{2}{c}{Negotiation}         & Measurement \\ \hline
\textbf{Alice}          & Tx              & Rx              & Tx       \\
\textbf{Bob}            & Rx              & Tx              & Rx       \\ \hline \hline
\textit{Sensing Attack} & \multicolumn{2}{c}{Eavesdropping} & Jamming  \\ \hline
\textbf{Eve}            & Rx              & Rx              & Tx   
\end{tabular}
\end{table}

\subsection{WLAN Sensing -- Detailed Operation}
\label{sec:WLAN_sensing_description}
In this section, we detail the WLAN sensing framework, based on
\cite{du21}\cite{ropitault23}. For the envisioned use cases, we refer to \cite{wlan_sensing_scenarios}. We describe the sensing stages, the frame structure, the transmitted sensing signal, the surveillance channel, and finally the received-side signal processing chain.

\subsubsection{Stages of WLAN Sensing}
The WLAN sensing uses the protocols initially implemented for multi-user multi-input multi-output in the 802.11ac amendment\cite{bejarano13}. In essence, this protocol allows a Tx to trigger channel sounding with explicit feedback from an Rx. To do so, the Tx (Alice) emits two packets called null data packet announcement (NDPA) and null data packet (NDP). Then, Rx (Bob) estimates the channel transfer function (CTF) from the NDP. Depending on the configuration, Bob can either send the estimated CTF back to Alice, or it can compute the RDM itself. 
The WLAN sensing framework is composed of five stages, summarized as follows: 
\begin{enumerate}
    \item \emph{Sensing session setup} is when Alice discovers potential responders like Bob. Alice and Bob estimate the distance between each other through round-trip time as explained in Appendix \ref{app:rtt}, where we assume that LOS is present and resolvable. 
    \item \emph{Sensing measurement setup} is when Alice and Bob agree on sensing parameters (such as bandwidth, number of antennas, carrier frequency, pulse repetition interval, and subcarrier grouping) with unprotected over-the-air transmission. Hence, any device can eavesdrop to deduce the sensing parameters.
    \item \emph{Sensing measurement instance} is when the sensing takes place, i.e., Alice emits NDPA/NDP, and Bob estimates the CTFs. \emph{WLAN sensing resembles a pulsed-OFDM radar scheme}, where successive NDPA/NDP transmissions are separated by a fixed pulse repetition interval. Since the standard defines the number of samples in NDPA/NDP, Bob knows exactly how many to collect before waiting for the next pulse. This will play a crucial role when the jammer framework is introduced.
    \item \emph{Sensing measurement and session termination} are the two stages where Alice and Bob release their resources dedicated to sensing. 
\end{enumerate}
For ease of reference, we refer to the combination of the sensing session setup and sensing measurement setup as sensing \emph{negotiation}.

\subsubsection{Frame Structure during Sensing Measurement}
\label{sec:frame_structure}
\begin{figure}
    \centering
    \includegraphics[width=0.8\linewidth]{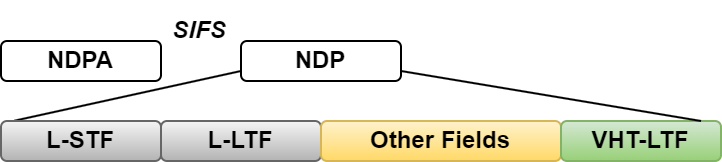}
    \caption{During a sensing measurement instance, Alice transmits an NDPA and an NDP, separated by SIFS seconds. Only the last field of the NDP, VHT-LTF, is used in channel estimation for sensing.}
    \label{fig:sensing_frame_structure}
\end{figure}

\review{Following the negotiation stages, Alice transmits two frames per sensing measurement instance. These frames are the NDPA and NDP as shown in \autoref{fig:sensing_frame_structure}.}
\review{The NDPA is a management frame, and its transmission announces that an NDP will follow. It contains information about which devices are involved in the sensing process and typically lasts around $40$ to $80$ $\mu s$. The NDPA triggers the Rx to prepare for receiving an NDP. }
\review{The NDP is a control frame, transmitted after the NDPA following a short inter-frame space (SISF), typically lasting 10 $\mu s$. It contains only the preamble and is used by Bob to perform channel estimation. Bob obtains a range profile with each NDP, and the Doppler profiles are computed over multiple NDPs. The NDP consists of different fields as shown in \autoref{fig:sensing_frame_structure}. The legacy short training field (L-STF) and legacy long training field (L-LTF) are used for joint time-frequency synchronization \cite{vandebeek99,schmidl97}. The other fields contain higher-level information. Finally, the very high throughput long training field (VHT-LTF) is used in channel estimation for sensing\footnote{We focus only on the 802.11ac amendment, which has the VHT acronym. Newer versions of the standard, such as 802.11ax, have different acronyms.}. Without losing any generality, and for clarity, we assume that each sensing measurement instance is composed of only the VHT-LTF, serving for both synchronization and channel estimation. This simplification allows us to model the synchronization and channel estimation more easily.}

\subsubsection{Transmitted Signal}
The OFDM parameters are defined as follows: $B$ is the signal bandwidth, $f_c$ is the carrier frequency, $Q$ is the number of subcarriers, $Q_{cp}$ is the cyclic prefix (CP) length, $M$ is the number of OFDM symbols, $q$ and $m$ represent the subcarrier and slow-time indices, respectively, $T=1/B$, $T_o$ and $T_s$ correspond to the sampling interval, the duration of an OFDM symbol including the CP, and the pulse repetition interval (PRI) which is an integer multiple of $T_o$, respectively and $T_s\gg T_o$, and $\Delta_f=1/(QT)$ is the subcarrier spacing. What follows in this section takes place during the sensing measurements instance, hence, we assume that the sensing negotiation already took place. 
The signal transmitted by Alice is then defined as $ \mat{S} \matsizer{Q}{M}$ in the frequency domain whose $M$ columns are identical. \review{Here, $\mat{S}$ contains standardized BPSK symbols making channel estimation quite straightforward with element-wise divisions\cite{horlin08}}. In the time domain, the signal structure resembles a pulsed radar where the OFDM symbols are separated by $T_s$ seconds as shown on \autoref{fig:scenario_topology}.

\subsubsection{Surveillance Channel} 
Let us define the steering vectors for the propagation delay as follows
\begin{align}
    \mat{d}(\tau) = [1\;e^{-j2\pi \tau\Delta_f} \; \hdots \; e^{-j2\pi \tau(Q-1)\Delta_f}]^T\matsizec{Q}{1},
\end{align}
where $\tau_p$ is the bi-static propagation delay. Similarly, the steering vector for the Doppler frequency shift is defined as
\begin{align}
    \mat{b}(f) = [1\;e^{-j2\pi f T_s} \; \hdots \; e^{-j2\pi f (M-1)T_s}]^T\matsizec{M}{1}, 
\end{align}
where $f$ is the Doppler frequency.
\review{The surveillance channel is modeled under the following assumptions: i) each object is a point in space with diffuse scattering characterized by its radar cross-section, and ii) each path can refer to any reflection, e.g., walls, furniture, mobile objects, etc. These generic assumptions are sufficient for this study \cite{durgin, wei24} since we focus on exposing the vulnerabilities at the signal processing level.} The channel model in the frequency domain is defined as
\begin{align}
    \mat{H} = \sum_{p=0}^P \alpha_p \mat{d}(\tau_p) \mat{b}^{\mathsf{H}}(f_p), \matsizec{Q}{M}.
    \label{eq:H_qm_explicit}
\end{align}
The path index $p=0$ models the LOS with complex amplitude $\alpha_0=a_0 e^{-j2\pi \tau_0 f_c}$, where $a_0$ represents path gain. The remaining indices model the different paths, each with a complex amplitude $\alpha_p$, a bi-static propagation delay $\tau_p$, and a bi-static Doppler frequency shift $f_p$. The propagation delays are assumed to be sorted in increasing order, i.e., ${\tau_0<\tau_1<\hdots<\tau_{P}}$. Hence ${\Delta_\tau=\tau_P-\tau_0}$ represents the delay spread of the channel. 

\subsubsection{Receiver Signal Processing Stage 1 -- Time-Frequency Synchronization}
\label{sec:ideal_receiver}

In a bistatic geometry, the receiver, Bob, needs time and frequency synchronization\cite{du21, richards10}. To do so, Bob uses an auto-correlation algorithm to estimate the time of arrival based on the detection of amplitude peaks. Hence, assuming that LOS is present and dominant, the signal propagated through it will be used for joint time-frequency synchronization. We define the strongest peak at the output of the lag-1 auto-correlation \cite{chiueh12, mahmood16} without losing any generality as follows 
\begin{align}
    \Xi[n_0] &= \abs{\alpha_0}^2 e^{-j2\pi \eta T_s}+z_0,
    \label{eq:Xi_n0}
\end{align}
where $z_0$ is the noise sample obtained after correlation, and $n_0 = (\tau_0+\delta_t)/T$ is the sample index of the LOS signal at the correlator's output with $\delta_t$ modeling the clock offset (see Appendix \ref{app:jtfs} for derivations)\footnote{Since we have simplified the frame structure used for sensing, the phase in \autoref{eq:Xi_n0} evolves with $T_s$ which should be replaced by $T_o$ to be fully standard compliant in a real-life setting.}. Since the phase of the peak is associated with the CFO, $\eta$, between Bob and Alice, Bob can detect the peak at $k=n_0$ for time synchronization\footnote{This approach allows the receiver to synchronize with the time-of-arrival, but neither $\delta_t$ nor $\tau_0$ can directly be estimated from $n_0$.} and use its phase for frequency synchronization, correcting the CFO via ${\hat{\eta} = {\angle \Xi[n_0]}/({2\pi T_s}})$. \textit{This synchronization step will be crucial when the jamming framework is introduced.}

Following the joint time and frequency synchronization, Bob performs the standard OFDM demodulation (removal of CP and FFT over each symbol), leading to the following received signal in the frequency domain:
\begin{align}
    \mat{R} = \mat{H}_0 \odot \mat{S} + \mat{Z}, 
\end{align}
where the entries in $\mat{Z}\matsizec{Q}{M}$ correspond to the noise samples in the frequency domain with zero mean and $\sigma^2$ variance. Here, the CTF perceived by Bob $\mat{H}_0$ is defined as
\begin{align}
    \mat{H}_0 = \sum_{p=0}^P \alpha_p \mat{d}(\tau_p-\tau_0) \mat{b}^{\mathsf{H}}(f_p). \label{eq:CFT-H0}
\end{align}
Here, we assume no time and frequency synchronization errors. Due to prior time synchronization, the propagation delays $\tau_p$ are now \emph{relative} to the direct path $\tau_0$ propagation delay. This is indicated by the zero-index on $\mat{H}_0$. Meanwhile, the CFO is compensated without errors; hence, the corresponding term does not appear on the estimated CTF. \review{If LOS is missing in the Alice-Bob channel, Bob would synchronize with the first non-LOS (NLOS) path, i.e., $l=1$, and the delay/Doppler terms in \eqref{eq:CFT-H0} would be relative to that path.}

\subsubsection{Receiver Signal Processing Stage 2 -- Radar Processing}
The CTF estimated by Bob is defined as follows 
\begin{align}
    \hat{\mat{H}}_0 = \mat{R} \hdiv \mat{S} = \mat{H}_0 + \mat{Z} \hdiv \mat{S}. \nonumber
\end{align}
Since the training fields, $\mat{S}$, are composed of BPSK symbols, the CTF estimation does not affect the noise variance, hence, no enhancement in the noise energy.
The RDM, ${\hat{\mat{Y}}\matsizec{Q}{M}}$, is obtained through inverse discrete Fourier transforms (IDFT) over $q$ and discrete Fourier transforms over $m$ (DFT) as follows
\begin{align}
    \hat{\mat{Y}} &= \mat{F}_Q \mat{\hat{H}}_0 \mat{F}_M^{\mathsf{H}},\matsizec{Q}{M}.
    \label{eq:Y_matrix}
\end{align}
where matrix $\hat{\mat{Y}}$ contains $P+1$ peaks. The peak at zero-range/zero-Doppler corresponds to the direct path used as the reference, while the remaining $P$ peaks correspond to the target echoes.
Once an RDM is obtained, a constant false-alarm rate (CFAR) detector separates the target echo peaks from noise peaks \cite{richards10}. Finally, we define the RDM processing operator $\hat{\mat{Y}} = \rdm{\hat{\mat{H}}}$ which encapsulates the range IDFT and Doppler DFTs as described in \eqref{eq:Y_matrix}.

\subsection{Vulnerabilities}
For future reference, we underline the following facts since they will play a crucial role in jamming. 
    First, the transmitted signals, $\mat{S}$, are standardized and known. This makes the target spoofing and deceptive jamming with SDRs much easier since signal reconstruction can be bypassed.
    Secondly, Bob synchronizes to the peak with the largest amplitude at the output of its correlator. This allows Eve to deceive Bob in different ways.

\section{Target Spoofing and Deceptive Jamming Framework}
\label{sec:deceptive_jammer}
\review{The framework focuses on generating artificial channel transfer functions (CTFs) to modulate the OFDM symbols. When these symbols are transmitted, they will alter Bob's perception of the true CTFs. Furthermore, we assume that Eve knows the type of signal processing employed by Bob, i.e., RDM-based. If Bob uses other methods for sensing, the attack may be less effective. However, Eve can adapt its jamming method for a variety of methods, which should be considered in future work.}

The jammer, Eve, has two main goals. The first one is to \emph{inject artificial targets} into Bob's RDM as in target spoofing. The second goal is to \emph{invalidate the surveillance signal} and the true target echoes that it contains as in deceptive jamming. If Bob employs other sensing methods, e.g., channel state information-based sensing, Eve's attacking method should be updated accordingly. To achieve its goals, we assume that Eve knows the sensing parameters\footnote{Either by eavesdropping during sensing negotiation or sensing measurement. The former is more straightforward since it only involves demodulating the appropriate fields in the preamble. The latter is more complicated since it requires the estimation of all the parameters directly from samples.}  ($Q, Q_{cp}, M, T, f_c, T_s$) established between Alice and Bob during the sensing negotiation. Moreover, the channel parameters for each bistatic geometry (Alice-Bob, Alice-Eve, and Eve-Bob) are assumed to be different.

\subsection{Time Alignment between the Surveillance and Jamming Signals at Bob}
\begin{figure}
    \centering
    \includegraphics[width=\linewidth]{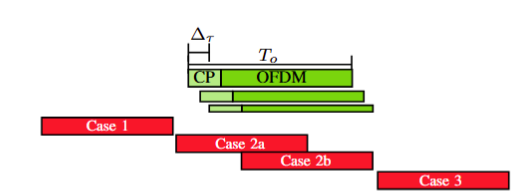}
    \caption{Different signal alignment cases during jamming. Green and red boxes correspond to Alice and Eve signals, respectively, while $\Delta_\tau$ and $T_o$ are the delay spread and OFDM symbol duration, respectively.}
    \label{fig:signal_alignments}
\end{figure}

Jamming takes place during a sensing measurement where Alice is transmitting sensing signals. \review{ Timing the signal transmission is crucial for Eve to ensure a successful attack. Although the distances, and the propagation delays, between the devices can be quite low in many scenarios, the frame structure shown in \autoref{fig:sensing_frame_structure} helps Eve in timing its signal transmission. Since an NDPA lasts at least $40\mu s$ \cite{du21}, Eve has enough time to detect the transmissions on air and time its transmission \cite{clancy11}.} \review{ Moreover, Bob has a time frame to receive these sensing signals. Hence, Eve's transmission should fall within this time frame. Since the duration of this time frame is not standardized, we consider various time alignment cases between Alice's and Eve's signals. Due to the nature of the OFDM waveform, these time alignment cases have different consequences.} The corresponding cases are shown in \autoref{fig:signal_alignments} and summarized as follows.
\begin{itemize}
    \item \emph{Case 1: Eve's signal arrives earlier than Alice's:} There is no alignment between the two signals\footnote{The tail of Eve's signal may align with the CP of Alice's signal. In this case, the surveillance channel may be observed at very large distances, e.g., hundreds of meters, and the peaks will be ignored by Bob.}. Hence, Bob synchronizes with Eve's signal, collects a predetermined amount of samples, and waits for the next OFDM symbol while completely omitting Alice's signal.
    \item \emph{Case 2: Alice's and Eve's signals are partially aligned:} Bob will detect two peaks at the correlator output and synchronize with the one having the largest amplitude \cite{chiueh12}. Whether Eve's signal arrives earlier (2a) or later (2b) has different consequences which are detailed in \autoref{sec:jammed_time_frequency_synchronization}.
    \item \emph{Case 3:  Eve's signal arrives later than Alice's:} Bob will only sample Alice's signals, causing jamming to fail.
\end{itemize}
Since the 802.11 standard does not dictate specific receiver design, hardware vendors have the flexibility to determine Bob’s timing window for signal reception. If Bob has a reliable RTT-based distance estimation and is configured to operate with a very narrow timing window, then the scenario described in Case 1 would likely be invalid. In such a configuration, Bob would only be vulnerable against Case 2.

\subsection{Jammer Operation}
The mathematical models are only derived for Case 2, which is the most general one. Due to the introduction of the jamming signal, there will be four different sets of channel parameters and CFOs, i.e., path gains, delays, Doppler shifts, and angles per channel. These are summarized in  \autoref{tab:channel_parameters} for easy reference. 
\begin{table}
\centering
\begin{tabular}{l|cccc}
Parameter\textbackslash{}Channel & A-B        & B-E              & E                \\ \hline
Number of targets                & $P$        & $L$              & 1            \\
Path index                       & $p$        & $l$              & -              \\
Path gain                        & $\alpha_p$ & $\alpha_l^{(b)}$ & $\bar{\alpha}$ \\
Path delay                       & $\tau_p$   & $\tau_l^{(b)}$   & $\bar{\tau}$   \\
Path Doppler frequency           & $f_p$      & $f_l^{(b)}$      & $\bar{f}$      \\
Path angles                      & $\theta_p$ & $\theta_l^{(b)}$   & -   \\
CFO                              & $\eta$     & $\eta^{(b)}$     & $\bar{\eta}$    
\end{tabular}
\caption{From left to right, the columns correspond to the Alice-Bob (A-B) and Bob-Eve (B-E) pairs, while the last column corresponds to the parameters of the artificial target generated by Eve (E).}
\label{tab:channel_parameters} 
\end{table}
Even though the number of targets can potentially be equal (i.e., $P=L$), their channel parameters will not be, i.e., amplitudes $\alpha_p\neq\alpha_l^{(b)}$, propagation delays $\tau_p\neq\tau_l^{(b)}$, and Doppler frequencies $f_p\neq f_l^{(b)}$ for $\forall p, l$.

\subsubsection{Jamming Transmitter}
Eve generates the deceptive signal by reversing the radar processing stages and exploiting the delay-frequency and time-Doppler dualities in \cite{durgin}. \review{Let us define the artificial RDM as $\bar{\mat{Y}}_j$ that contains two peaks: i) at zero-range/zero-Doppler with unit amplitude, corresponding to the reference peak, and ii) at $\bar{\tau}$ delay and $\bar{f}$ Doppler frequency with $\bar{\alpha}$ amplitude, corresponding to the artificial target. For the sake of simplicity, we include only one artificial target, but in more advanced scenarios, many artificial targets can be included. We also focus on generating an artificial RDM for a single radar snapshot. However, Eve can alter the artificial path parameters ($\bar{\alpha}, \bar{\tau}, \bar{f}$) over multiple snapshots, and generate multiple RDMs $\bar{\mat{Y}}_j$ successively. For a single snapshot, applying the radar processing in reverse order yields
\begin{align}
    \bar{\mat{H}} &= \mat{F}_Q^{\mathsf{H}}\bar{\mat{Y}}_j \mat{F}_M = \mat{1} + 
    \bar{a}\mat{d}(\bar{\tau})\mat{b}^{\mathsf{H}}(\bar{f}), \matsizec{Q}{M}\label{eq:H_bar}.
\end{align}
}
Here, $\mat{1}\matsizer{Q}{M}$ is a matrix of ones which allows us to generate a reference peak at zero-range/zero-Doppler. With the second term, a single artificial and mobile target is generated with amplitude $\Bar{\alpha}$, propagation delay $\bar{\tau}>0, \Bar{\tau}\in\mathbb{R}$, and Doppler frequency ${\Bar{f}\in\mathbb{R}}$. Then, after modulating each subcarrier with $\Bar{\mat{H}}$ and $\mat{S}$, the symbols that will be transmitted are defined as ${\mat{\bar{H}}\odot\mat{S}, \matsizec{Q}{M}}$. Finally, these modulated OFDM symbols can now be transmitted in the time domain by computing the IDFT over $q$ and adding the CP.

\subsubsection{Jamming Channel}
During the sensing measurement instance, Eve transmits the jamming signals through the channel between itself and Bob. We denote the corresponding frequency-domain channel by
\begin{align}
    \mat{B} &= \sum_{l=0}^L \alpha^{(b)}_l \mat{d}(\tau^{(b)}_l) \mat{b}^{\mathsf{H}}(f^{(b)}_l) \matsizec{Q}{M}.
    \label{eq:B_qm_explicit}
\end{align}
Here, $ \alpha^{(b)}_l$, $\tau^{(b)}_l$, and $f^{(b)}_l$ correspond to the amplitude, propagation delay, and Doppler frequency of the $l$th path between Eve and Bob. Meanwhile, ${l=0}$ refers to the LOS, and since the device is assumed to be stationary, $f^{(b)}_0=0$.

\subsubsection{Receiver Signal Processing Stage 1 -- Time-Frequency Synchronization}
\label{sec:jammed_time_frequency_synchronization}
If a jammer is present, Bob can be forced to synchronize with the jammer signal. The only requirement is that the signal transmitted by Eve, and propagated through the LOS with Bob (Eve-LOS), should have 3 dB or more power at Bob than the one transmitted by Alice, and propagated through the corresponding LOS (Alice-LOS)\cite{chiueh12}.

As described in \autoref{sec:ideal_receiver}, Bob computes the lag-1 auto-correlation over the received samples. In this case, there are two strict peaks at the following indices: ${k=n_0}$ and ${k=n_0^{(b)}}$ where ${n_0^{(b)}=(\tau_0^{(b)}+\bar{\delta}_t)/T}$ represents the time of arrival of Eve-LOS, with $\bar{\delta}_t$ being the clock offset. Similar to the definition of $\Xi[n_0]$ in \eqref{eq:Xi_n0}, $\Xi[n_0^{(b)}]$ can be defined as
\begin{align}
    \Xi[n_0^{(b)}] = \abs{\alpha_0^{(b)}}^2 e^{-j2\pi \eta^{(b)} T_s} + z_0, 
\end{align}
where $\eta^{(b)}$ refers to the CFO between Eve and Bob. Since Bob synchronizes to the peak with the largest magnitude at the output of the correlator, and if we assume that ${{20\log_{10}(\abs{\alpha^{(b)}_0}) -20\log_{10}(\abs{\alpha_0})>3\text{dB}}}$ (whether because ${d_{eb} < d_{ab}}$ or because Eve's transmit power is adjusted accordingly), Bob will time and frequency synchronize to Eve's jamming signal. This forced synchronization comes with different consequences.
\begin{itemize}
    \item \emph{Time synchronization implication:} If ${n_0^{(b)} < n_0}$, i.e., the jamming signal arrives \emph{earlier} than the surveillance signal, Bob will be time synchronized with an earlier clock corresponding to the Case 2a. Hence, the targets on the surveillance RDM will be range-shifted\footnote{In case target echoes arrive later than $n_0^{(b)}$, there will be also intersymbol interference. However, considering the minimum duration of the CP (which is $0.8\mu s$), the targets should be beyond 120 meters to satisfy the given condition. Hence, we omit this interference term.}. On the other hand, if ${n_0^{(b)} > n_0}$ the jammer signal arrives \emph{later} than the legitimate signal. If in addition ${n_0^{(b)} - n_0 > Q_{cp}}$ and $T_o=T_s$, the surveillance signal will be sampled beyond its CP, destroying its subcarrier orthogonality. These two cases have been studied in our earlier publication \cite{yildirim24} and also in \cite{lapan12, zhao19}.
    \item \emph{Frequency synchronization implication:} Bob will synchronize to the CFO of Eve, i.e., $\eta^{(b)}$. Hence, assuming the CFO is estimated without any errors, and if ${\eta - \eta^{(b)}}$ is significantly large\footnote{Usually above $B/(2Q)$ is sufficient.}, the subcarrier orthogonality of the surveillance signal will be completely lost, turning it into interference regardless of the time alignment between the two signals as studied in \cite{lapan13}. Hence, Bob will not be able to observe the surveillance channel at all. 
\end{itemize}

In summary, if Bob can be forced to synchronize with Eve, Alice's surveillance signal will experience inter-carrier interference (ICI), mainly due to the CFO, but potentially due to sampling beyond the CP. Regardless, the surveillance RDM will be corrupted. Moreover, the CFO will shift the surveillance RDM along the speed dimension since it has the same effect as the Doppler frequency shift.

Following the time-frequency synchronization, Bob perceives the channel between Eve and itself as follows
\begin{align}
    \mat{B}_0 &= \sum_{l=0}^L \alpha^{(b)}_l \mat{d}(\tau^{(b)}_l-\tau^{(b)}_0) \mat{b}^{\mathsf{H}}(f^{(b)}_l), \matsizec{Q}{M}
    \label{eq:B0_qm_explicit}
\end{align}
where the delays are modeled relative to the direct path.

\review{If LOS is missing in the Eve-Bob channel, the forced synchronization would still work as long as the first Non-LOS (NLOS) path, i.e., $l=1$, is stronger than the LOS (or NLOS) in the Alice-Bob channel. However, the drawback is that the delays and Dopplers in \eqref{eq:B0_qm_explicit} would become relative to the NLOS path. Hence, the artificial target will not appear on the intended RDM cell.}

\subsubsection{Receiver Signal Processing Stage 2 -- Radar Processing}
Assuming that Bob is force-synchronized with Eve, and demodulates the OFDM symbols, the estimated CTF takes the following form
\begin{align}
    & \hat{\mat{H}}_j = \underbrace{\mat{B}_0 \odot \bar{\mat{H}}}_{=\mat{G}_1} + \underbrace{\mat{H}'\odot\mat{C}}_{=\mat{G}_2} + \mat{Z}.
    \label{eq:bar_H_j_matrix}
\end{align}
Here, $\bar{\mat{H}}$ is the artificial CTF,  introduced in \eqref{eq:H_bar}, $\mat{B}_0$ corresponds to the physical channel between Eve and Bob, $\mat{H}'$ is the 
desynchronized surveillance channel, and $\mat{C}$ is the ICI. They are now described in detail. 
Hence, the first Hadamard product yields
\begin{align}
    \mat{G}_1\!\!=\!\!\mat{B}_0 \odot \bar{\mat{H}}
    &\!=\!\mat{B}_0\!+\!\sum_{l=0}^L \alpha^{(b)}_l\bar{\alpha} \mat{d}(\tau^{(b)}_l\!-\tau^{(b)}_0\!+\bar{\tau}) \mat{b}^{\mathsf{H}}(f^{(b)}_l\!+\bar{f}).
    \label{eq:first_hadamard}
\end{align}
Compared to the channel $\mat{H}_0$ observed in \eqref{eq:CFT-H0}, there are different types of targets observed in $\mat{G}_1$. The first term corresponds to the channel between Eve and Bob $\mat{B}_0$ and is present due to the reference peak in the artificial RDM. The second term with $l=0$ corresponds to the artificial target injected on delay/Doppler cell $(\bar{\tau}, \bar{f})$ when $l=0$. The remaining $L$ terms on delay/Doppler cells $(\tau_l^{(b)}+\bar{\tau}, f_l^{(b)}+\bar{f})$ are the real targets affected by the presence of the artificial target.

The desynchronized surveillance channel is
\begin{align}
    \mat{H}' = \sum_{p=0}^P \alpha_p \mat{d}(\tau_p-\tau') \mat{b}^{\mathsf{H}}(f^{(a)}_k+\eta_{w}),
\end{align}
where the delays are now relative to ${\tau'=\tau_0^{(b)} + \bar{\delta}_t}$ due to the time synchronization and ${\eta_{w}=\eta-\eta^{(b)}}$ is the combined CFO. Consequently, the desynchronized surveillance channel $\mat{H}'$ contains the same ${P+1}$ peaks as $\mat{H}_0$ but these peaks are shifted in delay by $\tau'$ and in Doppler by $\eta_w$.

Finally, the ICI, present due to the forced time and frequency synchronization, is modeled with $\mat{C} = \mat{P}\mat{S}\mat{\Lambda}$, 
\cite{horlin08}
where the entries in $\mat{P}\matsizec{Q}{Q}$ are defined as 
\begin{align}
    P[q,i] = \frac{ 1-e^{j2\pi (\frac{q-i}{Q} - \eta_wT)Q} }{ 1-e^{j2\pi (\frac{q-i}{Q} - \eta_wT)} } 
\end{align}
and the entries of the diagonal matrix ${\mat{\Lambda}\matsizec{M}{M}}$ are given as ${\Lambda[m,m]=e^{-j2\pi\eta_w mT_s}}$. For later reference, we define the second Hadamard product as ${\mat{G}_2 = \mat{H}'\odot \mat{C}}$. Here, $\mat{C}$ mixes the subcarriers and greatly affects the detectability of real target peaks.

The jammed and corrupted RDM is obtained by range/Doppler processing $\hat{\mat{H}}_j$ as follows
\begin{align}
     \hat{\mat{Y}}_j & = \fmat{Q}\hat{\mat{H}}_j\fmat{M}^{\mathsf{H}} \nonumber \\
    & = \underbrace{\fmat{Q} \mat{G}_1 \fmat{M}^{\mathsf{H}}}_{=\mat{Y}_{\mat{G}_1}} + \underbrace{\fmat{Q}\mat{G}_2\fmat{M}^{\mathsf{H}}}_{=\mat{Y}_{\mat{G}_2}} + \underbrace{\fmat{Q}\mat{Z}\fmat{M}^{\mathsf{H}}}_{=\mat{Y}_{\mat{Z}}}
    \label{eq:hat_Y_j_matrix}
\end{align}

\begin{remark}
    The model in \eqref{eq:hat_Y_j_matrix} can be generalized for the different time alignment cases from \autoref{fig:signal_alignments}:
\begin{align}
    \hat{\mat{Y}}_j=
    \begin{cases}
        \mat{Y}_{\mat{G}_1} + \mat{Y}_{\mat{Z}} &\text{Case 1} \\
        \mat{Y}_{\mat{G}_1} + \mat{Y}_{\mat{G}_2}+\mat{Y}_{\mat{Z}}&\text{Case 2} \\
        \hat{\mat{Y}} + \mat{Z} &\text{Case 3.} \\
    \end{cases}
    \label{eq:rdms_for_time_alignments}
\end{align}
If the time alignment corresponds to the first case, then only the first Hadamard product remains in \eqref{eq:hat_Y_j_matrix} since Alice's signal falls completely outside of the sampling window. The second case corresponds to the model provided in \eqref{eq:hat_Y_j_matrix}. Finally, the third case corresponds to the model provided in \eqref{eq:Y_matrix} where the surveillance channel is estimated as it is. Since Eve's jamming signals are not present at all, the jamming fails.
\end{remark}

\subsection{Achieving the Jammer Goals}
From \eqref{eq:hat_Y_j_matrix}, the RDM comprises two terms, given by the two Hadamard products, and they correspond to Eve's goals.

\subsubsection{Injecting artificial targets} 

This goal is achieved by the first Hadamard product, $\mat{Y}_{\mat{G}_1}$, which combines the RDM between Eve and Bob ($\mat{Y}_{\mat{B}_0}=\doubleFourier{Q}{\mat{B}_0}{M}$) with the artificially generated RDM ($\mat{Y}_{\bar{\mat{H}}}=\doubleFourier{Q}{\bar{\mat{H}}}{M}$). Here, $\mat{Y}_{\mat{B}_0}$ includes a reference peak at zero-range/zero-Doppler and physical targets at various range-Doppler cells, while $\mat{Y}_{\bar{\mat{H}}}$ includes an artificial mobile target and a reference peak. This product results in weighted, and range/Doppler-shifted copies of $\mat{Y}_{\bar{\mat{H}}}$ according to the peaks in $\mat{Y}_{\mat{B}_0}$, with its reference peak serving as the baseline.

\paragraph*{Limitations}
The main drawback of injecting artificial targets is that each peak in $\mat{Y}_{\mat{B}_0}$ adds an extra copy of $\mat{Y}_{\bar{\mat{H}}}$, potentially overcrowding the RDM. Although beneficial for Eve, this could alert Bob to an anomaly.

\subsubsection{Invalidating the surveillance RDM}
The second goal uses the Hadamard product $\mat{Y}_{\mat{G}_2}$, which reflects the interaction between the desynchronized surveillance channel $\mat{H}'$ and the ICI effects, $\mat{C}$. Without CFO and ICI, the surveillance channel remains observable as $\mat{Y}_{\mat{H}'}=\doubleFourier{Q}{\mat{H}'}{M}$. However, the ICI matrix $\mat{C}$ spreads the energy across the range dimension after the range IDFT. Then, the Doppler DFT focuses this energy along a single Doppler cell relative to CFO, causing a so-called ridge \cite{yildirim20}. Hence, each real target peak in $\mat{Y}_{\mat{H}'}$ will be replaced by range/Doppler shifted copies of this ridge.
\paragraph*{Limitations}
The first limitation is that Bob can detect the ridges, which are known indicators of OFDM-based radar interference and can respond effectively when detected. The second limitation is that Eve may unintentionally sabotage its artificial targets if they are aligned with the ridges.

\section{Advanced Jamming Strategies: A Qualitative Analysis}
\label{sec:jamming_strategies}

\begin{figure}
    \centering
    \includegraphics[width=\linewidth]{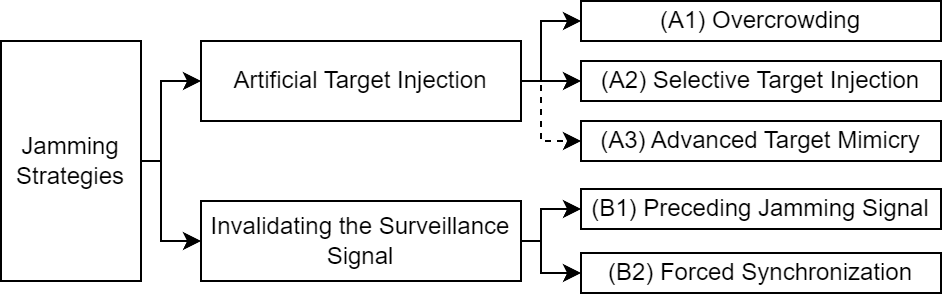}
    \caption{Different jamming strategies to achieve Eve's goals.}
    \label{fig:jamming_strategies_diagram}
\end{figure}

More advanced jamming strategies and their combinations are provided and discussed in this section, as visualized in \autoref{fig:jamming_strategies_diagram}. First, artificial target injection strategies will be discussed, where the options are either overcrowding (A1) or selective target injection (A2). Both of these strategies can be further enhanced with advanced target mimicry (A3). Second, we will discuss methods to invalidate the surveillance signal by preceding jamming signal (B1) or by forced synchronization (B2). Finally, we will qualitatively rate these strategies based on their simplicity/complexity and various effects on Bob.

\subsection{Injecting Artificial Targets}
\subsubsection{Overcrowding}
This type of target spoofing introduces \emph{many} targets to Bob, including artificial ones from Eve, real targets between Eve and Bob, and combinations of both. Hence, Eve cannot fully control where the real targets will appear in the range/Doppler cells. To overcrowd Bob's CFAR output, Eve keeps the first Hadamard product in $\eqref{eq:hat_Y_j_matrix}$ unchanged, introducing $2(L+1)$ targets. The artificial target appears at the intended RDM cell, while real targets are range/Doppler shifted based on the artificial target’s parameters.

\subsubsection{Selective Target Injection} 
This type of target spoofing gives Eve full control over the targets introduced to Bob. To ensure only the artificial target appears in Bob's CFAR output, Eve should exploit only the LOS path, which can be achieved by using multiple antennas. During the sensing negotiation, Eve estimates angles using subspace-based methods like MUSIC\cite{MUSIC} or ESPRIT\cite{ESPRIT}, then constructs a precoder to focus a beam towards Bob while nulling other real target paths. Once the radiation pattern is optimized, only the LOS term will remain in \eqref{eq:B0_qm_explicit}, and Bob will receive a single target at zero-range/zero-Doppler.

\subsubsection{Advanced Target Mimicry}
Bob may use technologies for target tracking and micro-Doppler analysis, and Eve can enhance deception by exploiting these. To force Bob to track the artificial target, Eve can update \eqref{eq:H_bar} over multiple snapshots following Newtonian kinematics, aligning with the tracking filters' equations. Simultaneously, Eve can mimic micro-Doppler signatures using empirical data or simulated patterns, such as simulating human walking motion with the Boulic model \cite{boulic90} via Matlab's radar toolbox.

\subsection{Invalidating the Surveillance Signal}
As pointed out in \eqref{eq:rdms_for_time_alignments}, Eve has two options to invalidate the surveillance signal. Eve can benefit from estimating the distances shown in \autoref{fig:scenario_topology}\footnote{The topology parameters, e.g., the LOS distances, can be known before jamming without signal processing. The attacker may access the network topology, e.g., networks in public areas, or perform other measurements, e.g., distance measurement with laser meters. Having access to such information makes the attacker more effective for jamming. However, for the sake of brevity, our analysis will only focus on estimating these parameters with signal processing.}. This can be done using the round-trip time when Alice and Bob alternate as Tx and Rx during the sensing negotiation stage, as outlined in Appendix \ref{app:rtt}. Although topology parameters, such as LOS distances, might be known beforehand, our analysis will focus on estimating them through signal processing.

\subsubsection{Preceding Jamming Signal}
If Eve can ensure the Case 1 signal alignment, then Bob will not perceive the surveillance signal. This corresponds to the best strategy for invalidating the surveillance signal. To successfully implement this strategy, Eve must accurately estimate the round-trip times (see in Appendix \ref{app:rtt}) and the transmission schedule of Alice. By doing so, Eve can time its transmissions so that Bob receives only the jamming signal, and collects the predetermined amount of samples, thereby rendering the legitimate signal invisible.

\subsubsection{Forced Synchronization}
If signals are aligned as in Case 2, Eve must force Bob to synchronize with itself rather than Alice. This leads to ICI on the surveillance signal, where ridges replace target peaks. Two conditions are necessary. First, Eve’s jamming signal power must exceed the surveillance signal power at Bob. To achieve this, Eve must know the distances between Alice and Bob ($d_{ab}$) and between itself and Bob ($d_{eb}$), and adjust its transmit power accordingly without overdoing it to remain stealthy. Second, Eve needs to estimate the CFO between itself and Alice ($\eta^{(a)}$) and Bob ($\eta^{(b)}$), then compute $\hat{\eta}=\eta^{(b)}-\eta^{(a)}$. Eve can introduce a much larger CFO, $\bar{\eta}$, ensuring a corrupted surveillance signal.

\begin{table*}[t]
\caption{\label{tab:strategy_rankings}Different jamming strategies. Requirements are defined as $\mathcal{S}$: single antenna, $\mathcal{M}$: multiple antennas, $\mathcal{P}$: additional processing unit for tracking/micro-Doppler signatures, $\mat{\Theta}$: LOS AoAs, $\mathcal{D}$: LOS distances, $\mathcal{T}$: timed transmission algorithm, $\mathcal{F}$: CFO estimation.}
\centering
\begin{tabular}{l|lllll}
Strategy                   & Requirement tags     & Complexity & Effectiveness & Detectability by Bob & Target presence \\ \hline
Overcrowding (A1)               & $\mathcal{S}$    & Very low   & Low  & High  & Artificial+True+Combined    \\
Selective Target Injection (A2) & $\mathcal{M},\mat{\Theta}$    & Moderate   & High& Moderate   & Artificial+True  \\
Advanced Target Mimicry  (A3)   & $\mathcal{P}$    & High  & Very High   & Low  & Artificial+True+Combined \\
\hline 
Preceding Jamming Signal (B1)  & $\mathcal{S/M,D,T}$    & High       & Very High   & Very Low  & Artificial+Combined  \\
Forced Synchronization (B2)    & $\mathcal{S/M,D,F}$    & Moderate   & High  & Moderate & Artificial+Ridges
\end{tabular}
\end{table*}
\subsection{Qualitative Analysis}
Now that various jamming strategies and different options to achieve them are discussed, they are qualitatively analyzed in \autoref{tab:strategy_rankings}.
    (A1) Overcrowding Bob is the simplest approach since it has no additional requirements other than a single antenna. However, it introduces the real targets along with the artificial ones. (A2) If only the artificial targets are desired at Bob, then multiple antennas are needed for beamforming. Then Eve can exploit the LOS with Bob, and avoid illuminating the real targets, greatly increasing the jamming effectiveness. However, it comes with increased complexity due to the AoA estimations and beamforming. (A3) Mimicking realistic target signatures is the most effective but complicated way to deceive Bob, especially when this is combined with A2. (B1) Preceding jamming signal provides the best method to invalidate the surveillance signal since it will not be observed at all. However, it requires the estimation of the device distances, transmission schedules, and a timed transmission algorithm which will greatly increase the complexity. (B2) Forced synchronization is a reliable option if B1 is not available, since ridges replace the true target peaks. However, the presence of the ridges can alert Bob, making it react accordingly. Moreover, Eve has to estimate the device distances to adjust its transmit power, and the relative CFO to create ICI.

\section{Numerical Analyses}
\label{sec:numerical_analyses}

\subsection{Simulation Parameters}
The radar parameters and topology information can be found in \autoref{tab:numerical_parameters}. We use the Blackman window for sidelobe suppression along the range and speed dimensions. It is important to note that the RDMs shown here are not derived from the numerical solution of \eqref{eq:Y_matrix} or \eqref{eq:hat_Y_j_matrix}. Instead, they result from a full radar chain simulation, including channel propagation with convolution, time-frequency synchronization with correlation, and subsequent radar processing.

\begin{table}
\caption{\label{tab:numerical_parameters} System and topology parameters for numerical analysis, where $p_A$, $p_B$, $p_E$, and $p_T$ correspond to the 2D coordinates of Alice, Bob, Eve and the Target, respectively. The velocity vector and radar cross-section of the target are indicated by $v_T$ and $\sigma_{T}$, respectively.}
\centering
\begin{tabular}{ll|ll}
Parameter    & Value   & Parameter    & Value   \\ \hline
$Q$          & 1024    & $p_A$        & $(10\meter, 0\meter)$  \\
$Q_{cp}$     & 64      & $p_B$        & $(0\meter, 0\meter)$   \\
$B$          & 80 MHz  & $p_E$        & $(5\meter, 10\meter)$  \\
$M$          & 128     & $p_T$        & $(5\meter, 10\meter)$  \\
$M_o$        & 100     & $v_T$        & $(-3\meter/\text{s}, -3\meter/\text{s})$ \\
$f_c$        & 5 GHz   & $\sigma_{T}$ & $0.1 \meter^2$      \\
\end{tabular}
\end{table}

\begin{figure}[h!]
    \centering
    \includegraphics[width=\linewidth]{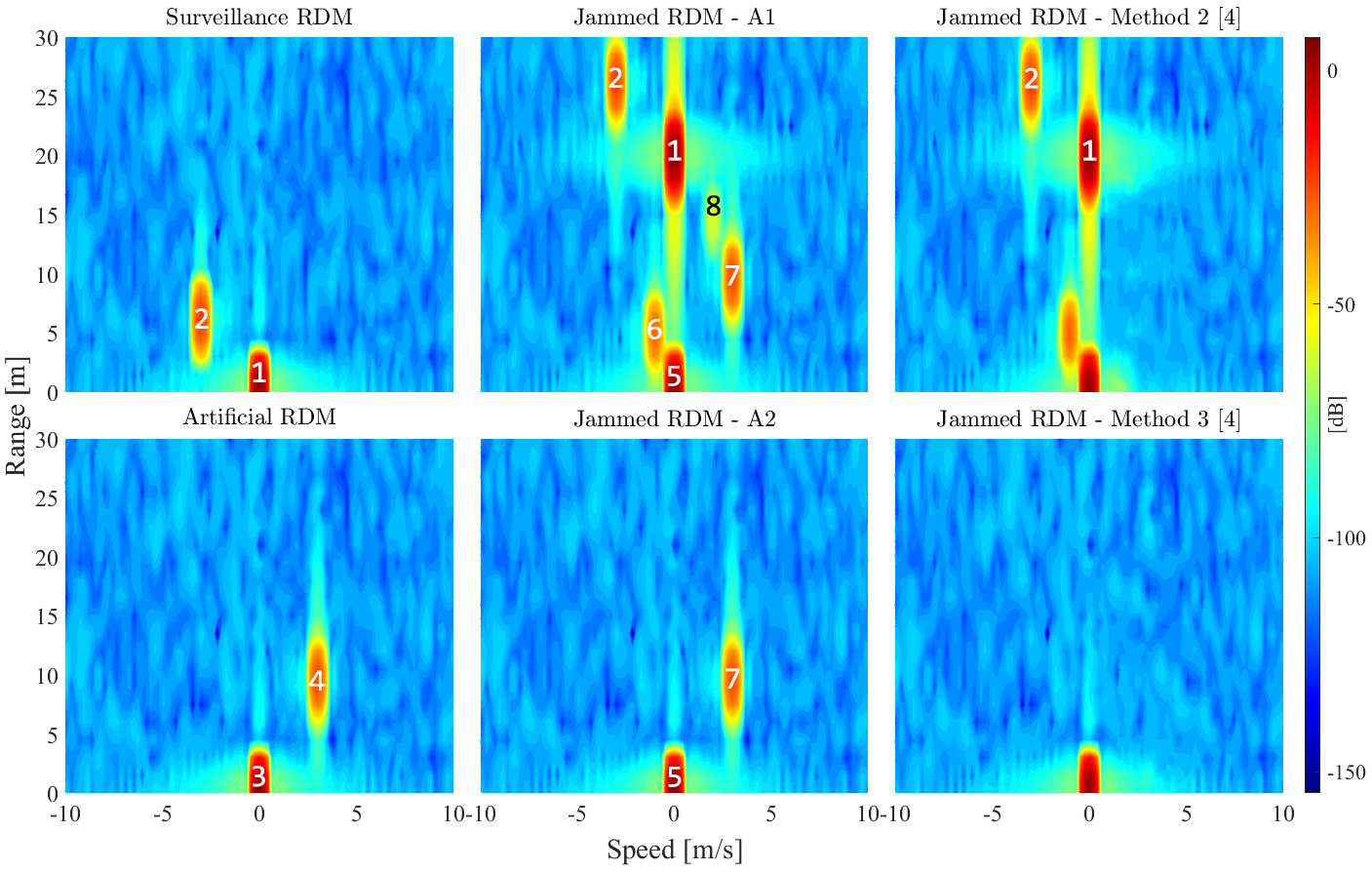}
    \caption{\review{Six RDMs are provided. The first column shows the surveillance and artificial RDMs in isolation. The second column corresponds to jammed RDMs with A1 and A2 strategies. The third column corresponds to the methods in \cite{zhang21b}, adapted and implemented for sensing. As a comparison, surveillance RDM and artificial RDM correspond to case 3 and case 1 types of time alignments, respectively.}}
    \label{fig:injecting_artificial_targets}
\end{figure}

\subsection{Results and Discussion} 
\begin{figure}[h!]
    \centering
    \includegraphics[width=\linewidth]{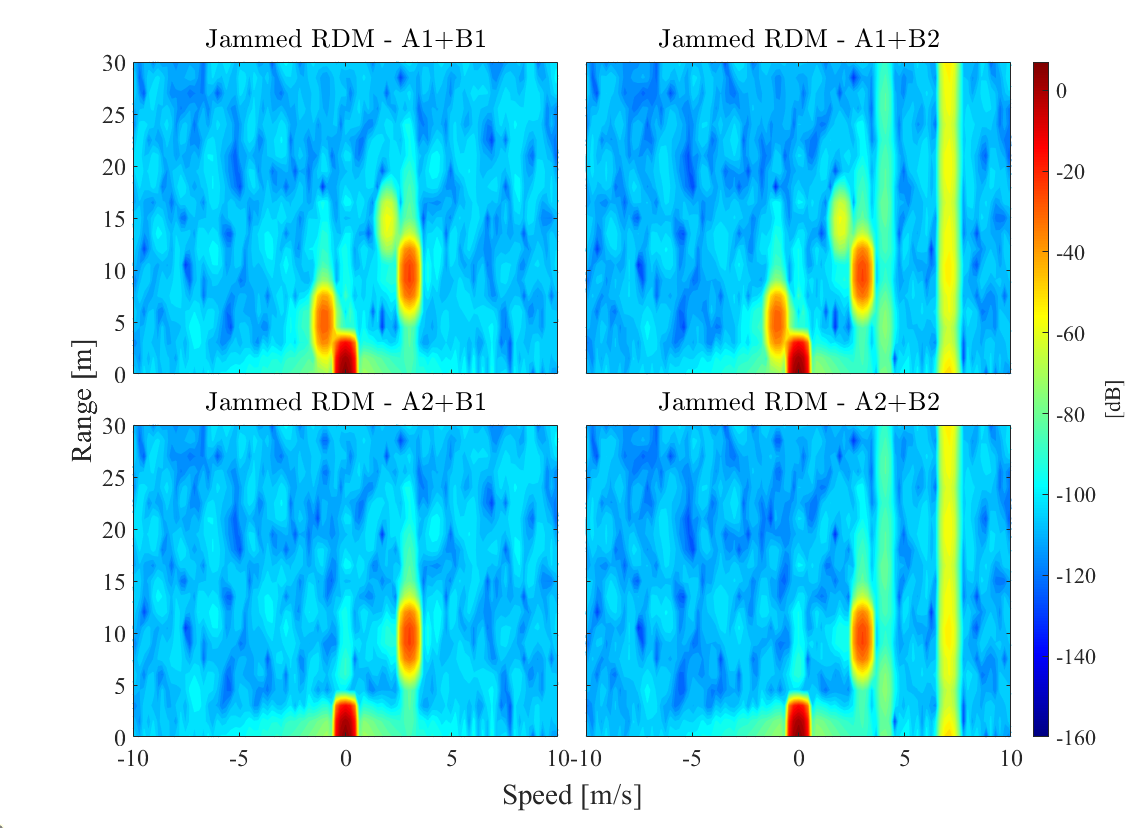}
    \caption{Four RDMs for different strategy combinations. The best jamming performance is the combination of A2 and B1 where only the artificial target peak is present.}
    \label{fig:combined_jamming_strategies}
\end{figure}

In this section, we provide results for the target spoofing and deceptive jamming performance of Eve. Initially, we analyze Bob's RDMs on a realization basis. In \autoref{fig:injecting_artificial_targets}, target spoofing strategies, and in \autoref{fig:combined_jamming_strategies} combined strategies are evaluated. Then, we evaluate the overall spoofing and jamming performance of Eve where the main key performance indicator is the probability of detection (PD) of targets. In \autoref{fig:pd_vs_cfo}, the PD is analyzed as a function of the artificial CFO. In \autoref{fig:pd_of_artificial_vs_real_targets}, the PD is analyzed as a function of jammer-to-signal-ratio (JSR). In \autoref{fig:mdr_vs_dr}, overall jamming performance is studied where we compare the missed detection rate of real targets with the PD of artificial targets. Finally, in \autoref{fig:number_of_detected_targets}, the detection rate is evaluated for the different probability of false alarms (PFa) and number of real targets. In all the detection-related studies, the OS-CFAR \cite{blake88} algorithm is used.

\subsubsection{Injecting Artificial Targets}
In \autoref{fig:injecting_artificial_targets}, the RDMs illustrate WLAN sensing and artificial target injection. The surveillance RDM shows the physical channel between Alice and Bob with synchronization peak \peak{1} and the target \peak{2}. Eve's artificial RDM contains synchronization peak \peak{3} and artificial target \peak{4}. The upper-middle RDM, A1 jamming, shows Bob synchronizing with Eve's signal, where peak \peak{5} corresponds to \peak{3}, and the real target between Eve and Bob appears as peak \peak{6}. Though at a different range/speed than \peak{2}, it is still detectable. The artificial target peak \peak{7} appears at its intended spot and the combination of artificial and real targets forms peak \peak{8}. Here, Eve's LOS arrives earlier than Alice's, shifting Bob’s time synchronization, and causing the surveillance RDM to appear beyond 20 meters. In other cases, peaks may be beyond CFAR detection limits, causing Bob to discard them. The bottom-middle RDM, A2 jamming, shows beamforming's effect—nulling the real target makes it undetectable, leaving only the artificial target visible. Here, Alice's LOS arrival time is intentionally forced to arrive much later during simulations, making the surveillance RDM disappear. These findings align with those in \cite{argyriou23}. \review{ For comparison, the throughput jamming methods from \cite{zhang21b} are adapted to disrupt the sensing session. In method 2, an injected preamble arrives before Alice's signal, which Bob uses to synchronize, causing a range shift on the surveillance RDM. However, the true target still appears. In method 3, the injected preamble arrives significantly later than Alice's signal, and due to the beamforming, no additional paths are illuminated. As a result, only the reference peak appears on the RDM. In both cases, Eve does not introduce artificial targets. These findings indicate that the methods in \cite{zhang21b} provide some effectiveness for deceptive jamming, but they are inadequate for target spoofing.}

\subsubsection{Combined Strategies}
In \autoref{fig:combined_jamming_strategies}, four RDMs display the effects of combining different jamming strategies. The top-left RDM shows overcrowding and preceding jamming, the true surveillance channel is obscured, but true and combined targets remain visible. In the top-right RDM overcrowding and forced synchronization are combined. Alice's and Eve's signals are somewhat aligned, and the surveillance RDM is present. However, ICI caused by CFO spreads the energy along the range dimension, forming ridges \peak{1} and \peak{2}. The speed dimension remains unaffected since symbol-to-symbol phase shifts remain unaffected, though Doppler shifts occur due to CFO. The bottom-left RDM combines selective injection and preceding jamming. It is the most effective strategy since only the artificial RDM is visible without extra targets or ridges. Lastly, the bottom-right RDM combines selective injection and forced synchronization, where surveillance peaks are replaced with ridges, though additional targets are absent due to beamforming.

\begin{figure}
    \centering
    \includegraphics[width=\linewidth]{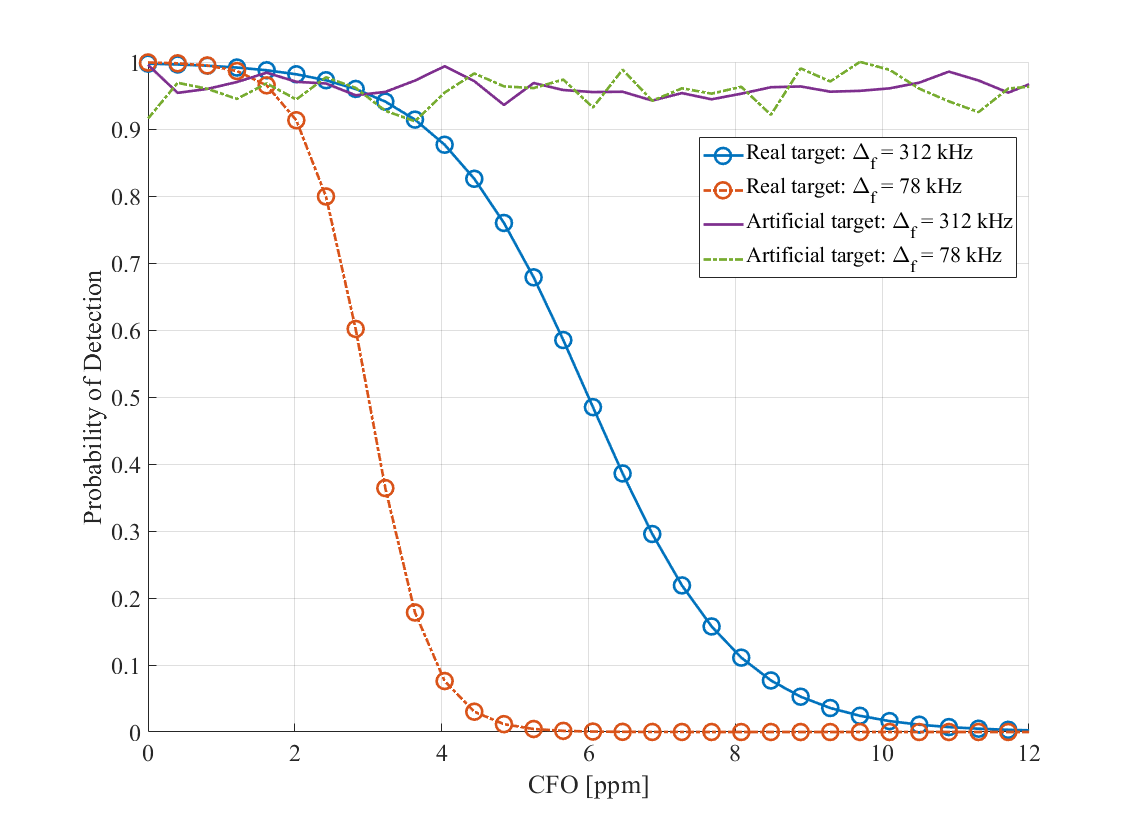}
    \caption{Probability of detecting real and artificial targets as a function of CFO for two different subcarrier spacings: 312kHz for 802.11ac and earlier, 78kHz for 802.11ax and later. The OS-CFAR \cite{blake88} algorithm is used for target detection, with $10^{-6}$ as the probability of false alarm. The real target speed is randomized over 5k realizations for each CFO value.}
    \label{fig:pd_vs_cfo}
\end{figure}

\subsubsection{Target Detection Probability vs. CFO}
In \autoref{fig:pd_vs_cfo}, the effect of CFO on the PD of real and artificial targets in the forced synchronization strategy is examined for 10 dB JSR. With wider subcarrier spacing (312 kHz, as in 802.11ac), when the CFO is below 3 ppm, the range IDFT focuses energy on a range cell, allowing the real target to be detected. However, above 3 ppm, \emph{the PD of the real target drops significantly} as energy spreads along the range, creating ridges. For narrower subcarrier spacing (78 kHz, as in 802.11ax), the system is more sensitive to CFO, with a critical threshold of around 1 ppm, beyond which PD rapidly declines.
In contrast, \emph{the PD of detecting the artificial target remains unaffected by subcarrier spacing or CFO}, as the CFO impacts only the surveillance signal, not the jamming signal. However, if the ridges caused by the CFO and the artificial target are aligned along the speed dimension, the artificial target may not be detected. Thus, Eve could unintentionally undermine its target spoofing. This explains why the PD of the artificial target is not 100\% and varies with different CFO values--sometimes the ridges conceal the artificial target. This is unavoidable since Eve does not know the real target's speed relative to the Alice-Bob channel.

\subsubsection{Target Detection Probability vs. Jammer-to-Signal-Ratio}
\begin{figure}
    \centering
    \includegraphics[width=\linewidth]{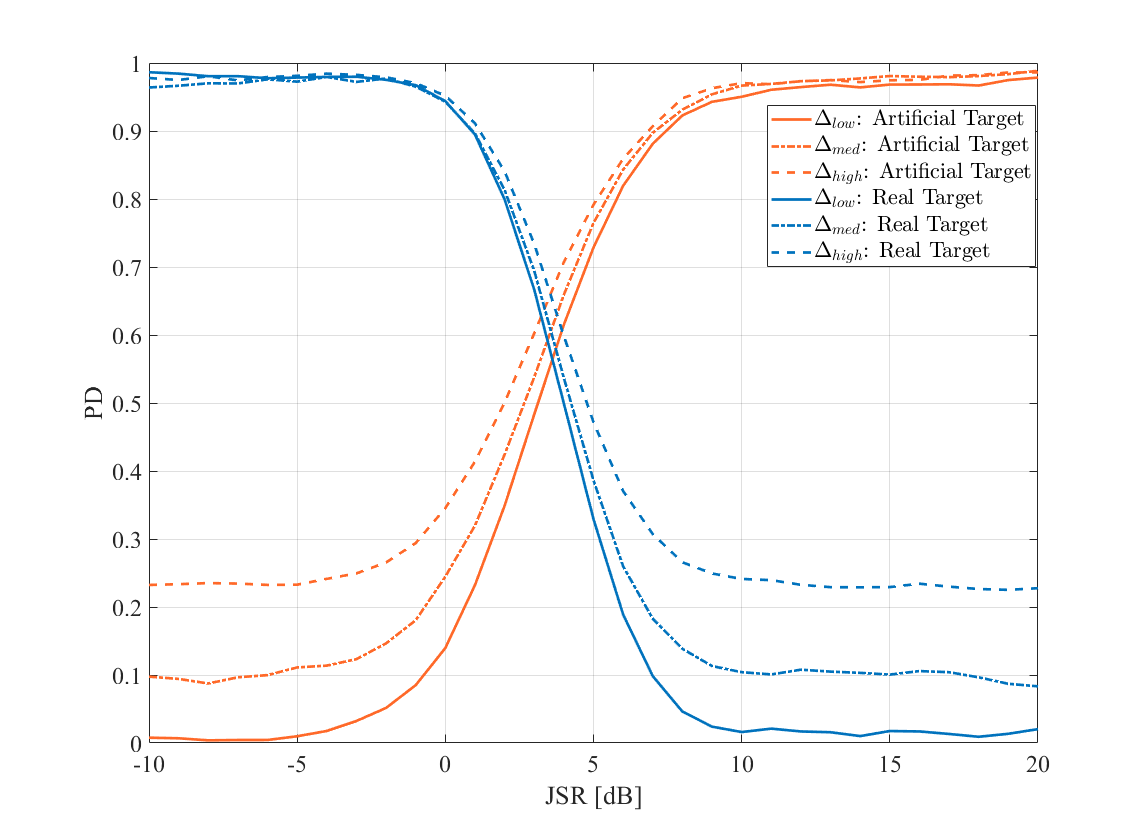}
    \caption{\review{The PD of artificial and real targets are shown as a function of JSR for three different CFO regions. The CFO difference between Eve and Alice is given as ${\Delta = \lvert \eta - \bar{\eta} \rvert}$ and the corresponding regions are defined as ${\Delta_{high} > 4\text{ ppm}}$, ${4 \text{ ppm } \geq \Delta_{med} > 1 \text{ ppm}}$,  and ${1 \text{ ppm} \geq \Delta_{low}}$. Above 8-10 dB JSR, the jamming works since Bob synchronizes to Eve, and Alice's signal turns into noise. Below this threshold, jamming does not work. }}
    \label{fig:pd_of_artificial_vs_real_targets}
\end{figure}

\review{In \autoref{fig:pd_of_artificial_vs_real_targets}, the effect of varying Eve's signal power relative to Alice's is analyzed by plotting the PD of both artificial and real targets as a function of the JSR across three CFO regions. The noise floor remains constant, ensuring that at 0 dB JSR, the jammer-to-noise ratio and SNR are both 30 dB. The PD values are derived from 10k realizations for each JSR and CFO combination, with the CFO randomly chosen within the defined regions. The time alignment between Alice's and Eve's signals also varies randomly between -24 and +24 samples, ensuring that CFO is the only source of orthogonality loss.}
\review{At low JSR values, the jamming does not work since Bob synchronizes to Alice's signal. This leads to a loss of orthogonality in Eve's signal, which the CFO influences. As a result, artificial target peaks in the RDM either i) stretch into ridges, reducing PD for high CFO values or ii) remain detectable for lower CFO values. During this phase, the PD of real targets is high as expected. As JSR increases, there is a transition in PD between real and artificial targets. When JSR exceeds 10 dB, the PD of artificial targets becomes high enough to ensure effective jamming. Meanwhile, real targets are increasingly missed due to the CFO. \emph{This region highlights strong target spoofing and deceptive jamming performances--indicated by the high PD of artificial targets and low PD of real targets, respectively.}}

\subsubsection{Missed Detection Rate vs. Detection Rate}

\begin{figure}
    \centering
    \includegraphics[width=\linewidth]{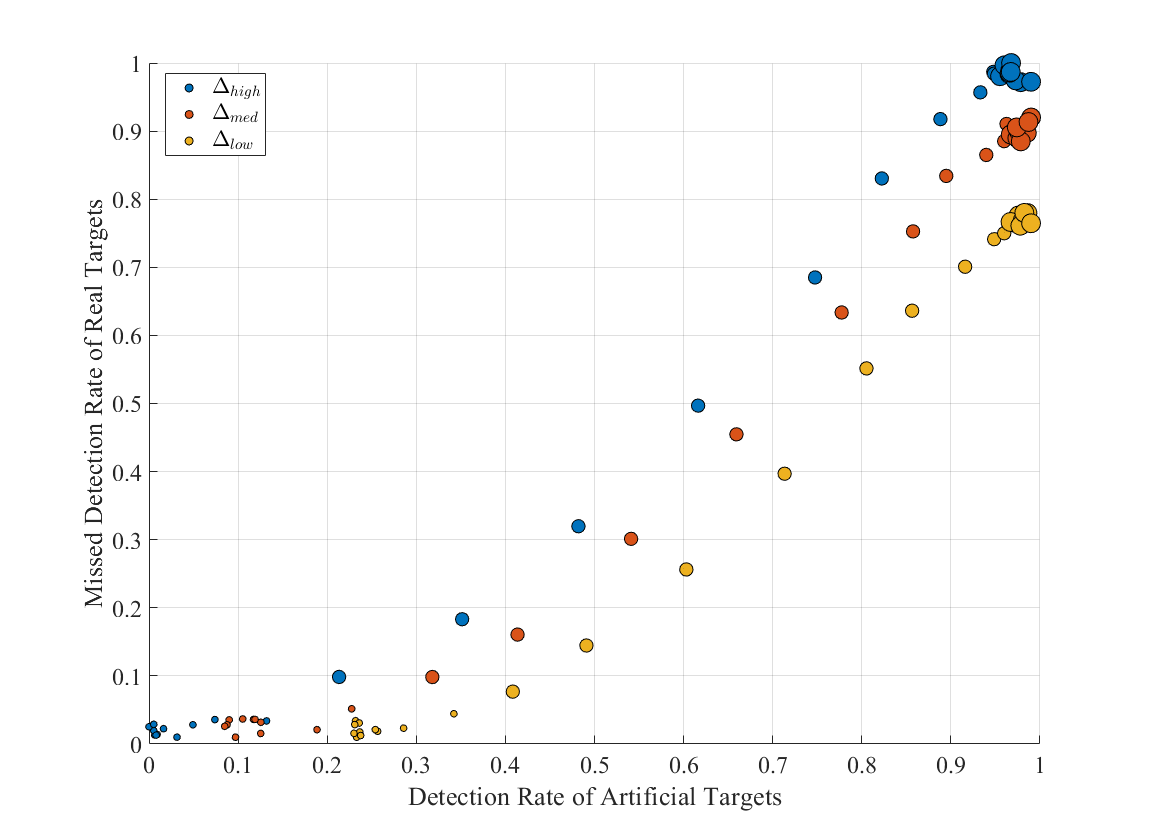}
    \caption{ \review{The MDRrt and DRat are plotted for three different CFO regions. Based on \autoref{fig:pd_of_artificial_vs_real_targets}, three JSR regions can be identified: low (below 0 dB), medium (between 0 and 10 dB), and high (above 10 dB). The size of the scatter plots are linked with these regions, e.g., high JSR points have the largest scatter points. The CFO difference between Eve and Alice is given as $\Delta = \lvert \eta - \bar{\eta}\rvert$, and the regions defined for \autoref{fig:pd_of_artificial_vs_real_targets} apply here.}}
    \label{fig:mdr_vs_dr}
\end{figure}

\review{\autoref{fig:mdr_vs_dr} further shows the effectiveness of the jammer by plotting the miss detection rate of real targets (MDRrt, obtained from the corresponding PD) and the detection rate of artificial targets (DRat) as a function of JSR.}
\review{Based on the plot, three JSR regions can be identified: low (below 0 dB), medium (between 0 and 10 dB), and high (above 10 dB). When JSR is low, the MDRrt is also very low. Meanwhile, for relatively lower CFO differences, the DRat reaches roughly 20\%. The medium JSR levels serve as a transition region where minimal differences in JSR significantly increase the MDRrt and DRat. As the system reaches high JSR levels, MDRrt and DRat reach their maximal values. Similar to the low JSR region, the real targets may be detected in some realizations when the CFO difference is low. \textit{Hence, Eve should operate above 10-12 dB JSR and with a high CFO difference for the best jamming performance.}}

\subsubsection{Overcrowding Detection Performance}
\begin{figure}
    \centering
    \includegraphics[width=\linewidth]{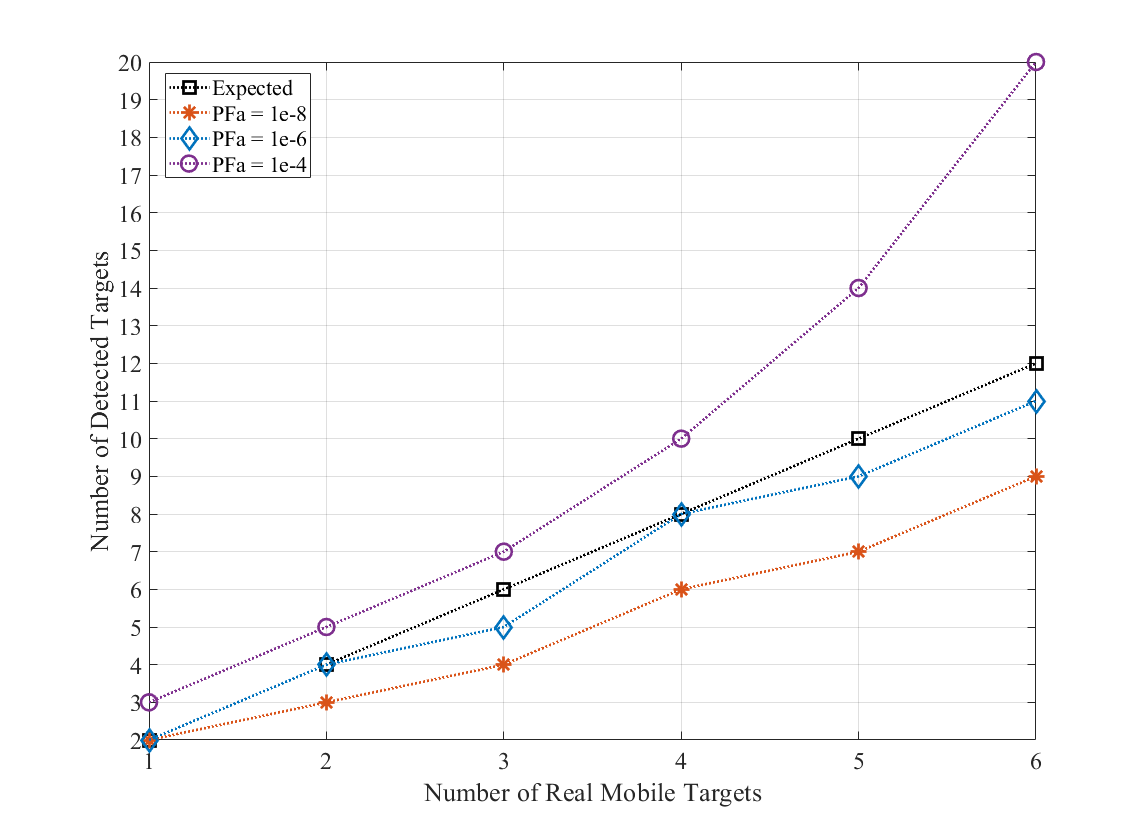}
    \caption{\review{The impact of the number of real scatterers in the environment is studied as a function of Bob's PFa. The JSR is fixed to 10 dB, and the SNR is 30 dB.}}
    \label{fig:number_of_detected_targets}
\end{figure}

\review{\autoref{fig:number_of_detected_targets} illustrates the number of detected targets as a function of the total number of targets in the environment, across three different false alarm rates (PFa) For each PFa and number of real targets, 10k realizations were simulated. In these simulations, Eve transmits an RDM with two peaks: one representing the artificial target and the other as a reference, accounting only for first-order reflections between these peaks and the targets. The expected number of detected targets is calculated using the analytical expression in \eqref{eq:first_hadamard}.}
\review{When PFa is $10^{-4}$, the detection algorithm is more permissive, leading Bob to detect more peaks than expected, i.e., some noise peaks are mistakenly identified as targets. At $\text{PFa}=10^{-6}$, the detection accuracy aligns closely with the expected values, though slight deviations still occur. Finally, at $\text{PFa}=10^{-8}$, the detection algorithm becomes overly strict, causing Bob to miss many real targets, which decreases the reliability of the sensing system. \textit{In essence, the overcrowding strategy works.} However, assuming that Eve does not have access to PFa used by Bob, the outcome is not entirely under the control of Eve.}

\section{Experimental Validation}
\label{sec:experimental_validation}
In this section, we explore the design and implementation of an SDR-based jammer by utilizing the USRP X310 platform.
\subsection{Experimental Setup}
\begin{figure} 
    \centering
    \includegraphics[width=\linewidth]{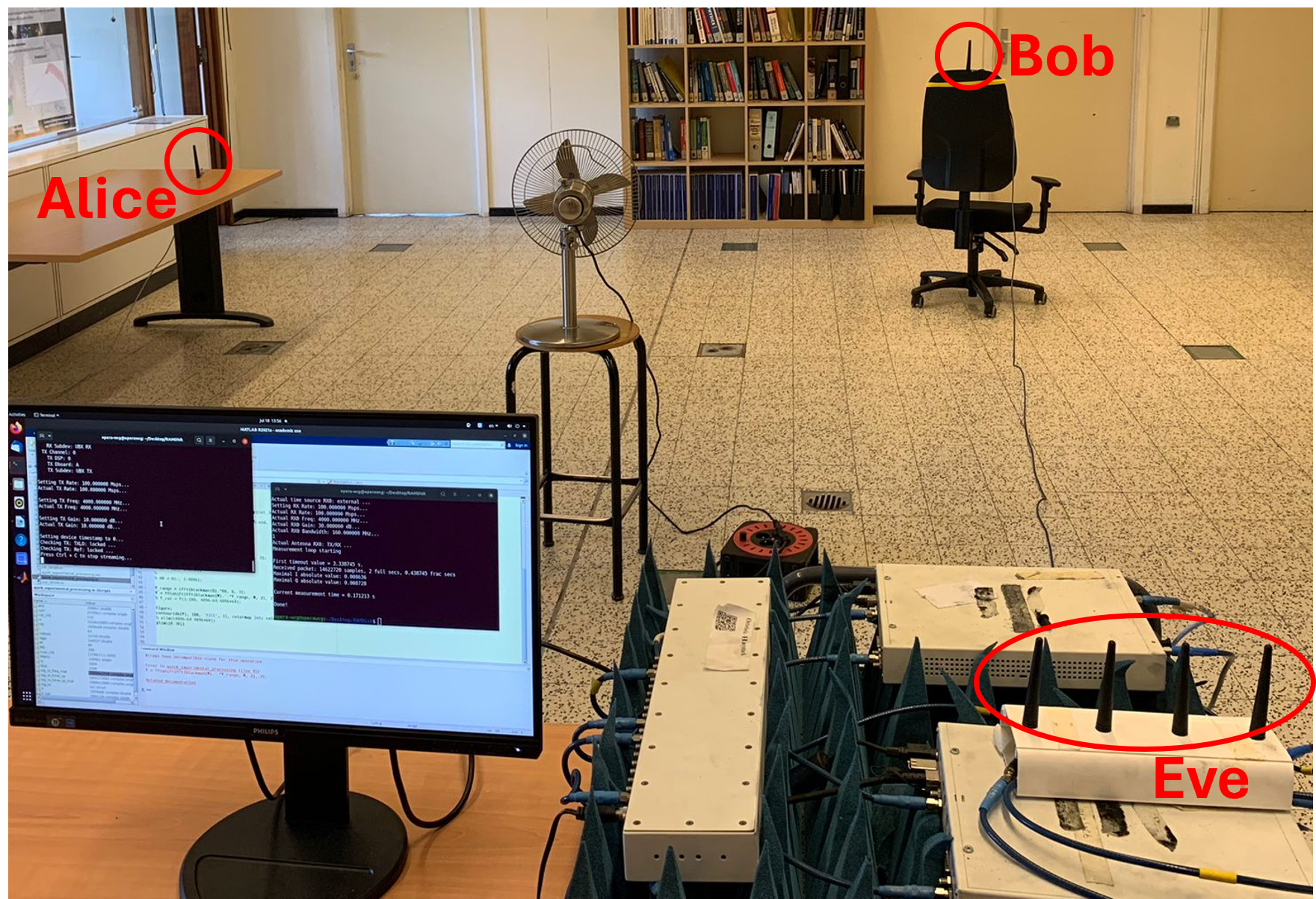}
    \caption{The picture of the experimental setup, consisting of a metallic fan as the mobile target and three USRP X310s: one to emulate Alice and Bob, and the remaining two to emulate Eve's array. For Eve's single antenna transmission: i) only the antenna at the far left is kept, and ii) the remaining antennas are removed to avoid coupling effects. For Eve's multiple antenna transmission: i) the AoA estimation stage is bypassed, instead, the angles are computed manually, and ii) the precoder matrix is designed to steer a null towards the target and a beam towards Bob}
    \label{fig:jamming_experiment}
\end{figure}

Our experimental setup is indoors and it is similar to the topology provided in \autoref{fig:scenario_topology}, as shown in \autoref{fig:jamming_experiment}. A metallic fan emulates a mobile target due to the consistent speed pattern obtained from its rotating blades, which helps diagnose RDMs. \review{In our models, we simplified the sensing frame structure by transmitting a single OFDM symbol used both for synchronization and channel estimation purposes. However, as described in \autoref{sec:frame_structure}, NDPA/NDP frames consist of multiple fields, such as L-STF and L-LTF for synchronization, and VHT-LTF for channel estimation \cite{du21}. In a real WLAN sensing setting, Eve must mimic L-STF and L-LTF to ensure that Bob synchronizes with Eve. However, we omit these fields in our experiments since i) they are straightforward to implement; and ii) they are unnecessary for our purposes, as VHT-LTF is the primary field for sensing processing.}

\begin{figure}
    \centering
    \includegraphics[width=\linewidth]{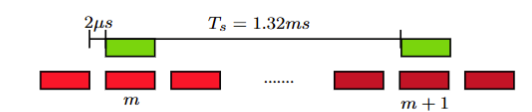}
    \caption{Signal structure for jamming. Alice transmits the same OFDM symbol with PRI $T_s=1.32 ms$, indicated with green. Meanwhile, Eve transmits in batches, each consisting of three identical OFDM symbols for time instant $m$, indicated with red.}
    \label{fig:jamming_signal_structure}
\end{figure}

The transmitted signals follow the structure shown in \autoref{fig:jamming_signal_structure}. Alice continuously transmits legitimate sensing signals, separated by $T_s = 1.32~\text{ms}$. Simultaneously, Eve transmits clusters of modulated OFDM symbols, with each cluster containing three identical symbols for the time instant $m$. This clustered transmission serves two purposes: (i) ensuring coverage for each time alignment case in \eqref{eq:rdms_for_time_alignments}, and (ii) increasing the probability of at least one jamming signal aligning with Alice's legitimate signal. The jamming signal clusters are also separated by $T_s = 1.32~\text{ms}$, using the middle symbol of each cluster as a reference.

\subsection{Results and Discussion}
We show the experimental surveillance channels between Alice and Bob and Eve and Bob. Then we show the impact of the different combined strategies (A1-B1, A1-B2, A2-B1, and A2-B2), to relate to the simulation results in \autoref{sec:numerical_analyses}.

\subsubsection{Surveillance Channel}
We present the RDMs for both the Alice-Bob and Eve-Bob channels, as shown in \autoref{fig:exp_onlyRealTarget}. In real-world wireless channels, static environmental features like walls and furniture generate multipath components (MPCs), appearing as clutter peaks around zero-Doppler in both RDMs. Four distinct peaks appear at approximately 3 and 6 meters in the bistatic ranges for the Alice-Bob and Eve-Bob channels, respectively. These peaks result from Doppler shifts caused by the fan’s blades. Additionally, ghost targets appear at farther ranges due to multipath effects. The clutter and fan blade peaks differ in each RDM due to the distinct bistatic geometries.

\subsubsection{Combined Strategies}
\begin{figure}
    \centering
    \includegraphics[width=\linewidth]{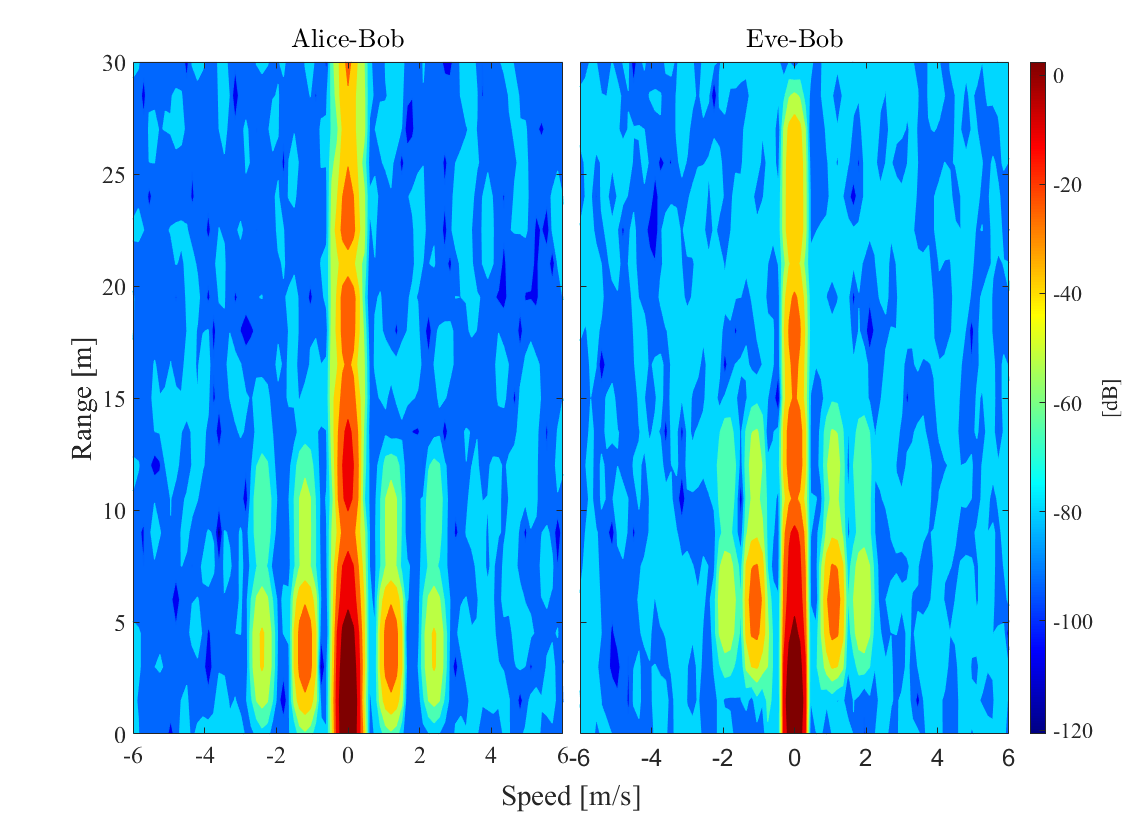}
    \caption{The RDMs obtained for Alice-Bob and Eve-Bob channels while the mobile target is active. Black stars mark the peaks in Alice-Bob RDM for comparison. \review{The environment appears as a set of static targets centered at 0m/s, and the fan's blades appear as four distinct peaks.}}
    \label{fig:exp_onlyRealTarget}
\end{figure}

\begin{figure}
    \centering
    \includegraphics[width=\linewidth]{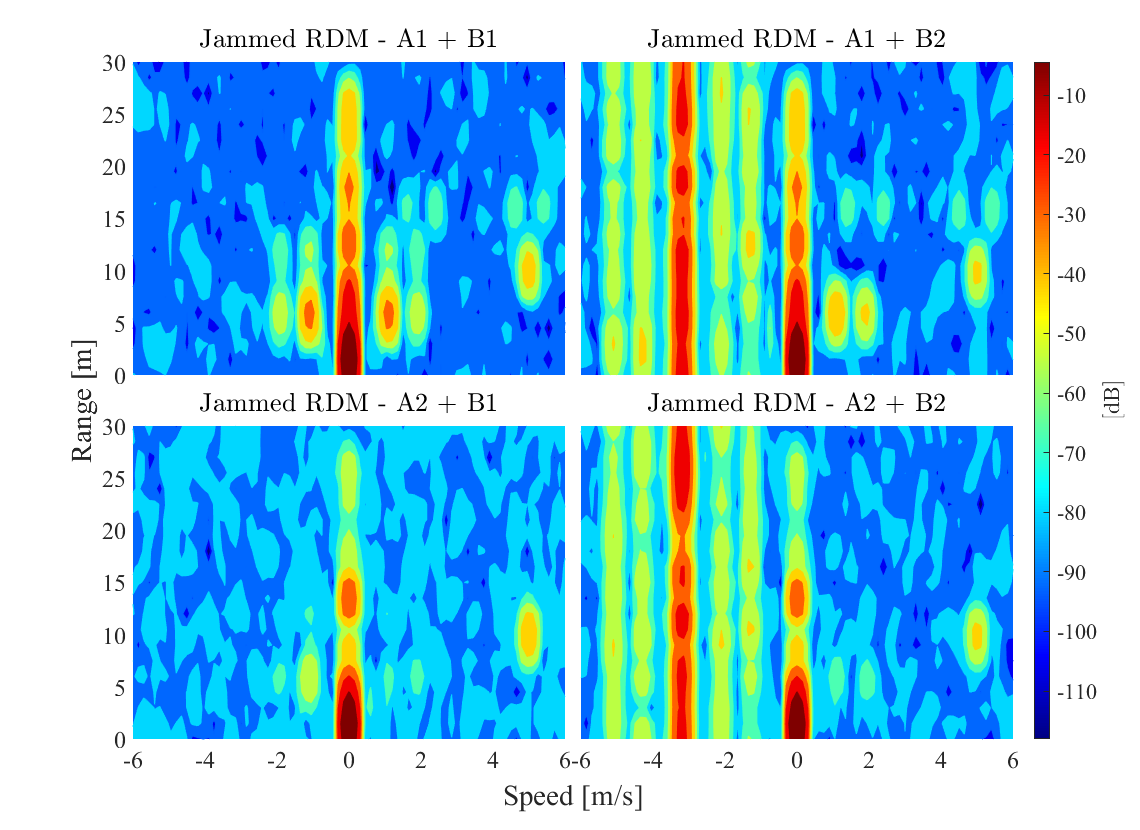}
    \caption{The jammed RDMs with combined strategies, with $\eta_w=5$ppm. The black stars correspond to the range/Doppler cells of the true target, while the red cross corresponds to the (10m,5m/s) cell for the artificial target. The ridges are caused by the loss of orthogonality on Alice's signals.}
    \label{fig:exp_combinedStrategies}
\end{figure}

We present four RDMs in \autoref{fig:exp_combinedStrategies}, each reflecting different combined jamming strategies. These RDMs include an artificial target at a 10-meter range and a speed of 5 m/s, marked with a red cross. The top-left RDM shows the first strategy, where the fan's signatures appear as peaks \peak{1}, similar to those in \autoref{fig:exp_onlyRealTarget}. The target peaks \peak{2} result from the interaction between the fan's signatures and the artificial target peak \peak{\times}. Additionally, Eve’s transmission of range/Doppler shifted signals for the artificial target creates a range/Doppler shifted clutter \peak{3}, scaled by $\bar{\alpha}$, enhancing the realism of the deception. The top-right RDM displays the effect of forced synchronization, where target peaks and clutter are replaced by ridges \peak{4} that shift along the speed dimension due to CFO, potentially folding over multiple times. The bottom-left RDM shows the most effective strategy, where Eve utilizes the directivity of a linear array. However, despite this, the mobile target’s signatures \peak{5} remain due to calibration errors and array sidelobes. The clutter pattern also differs from the top two RDMs because the scene is not isotropically illuminated. The bottom-right RDM, representing the last strategy, is characterized by ridges caused by the CFO.

\subsubsection{Comparison of Numerical and Experimental Results}
We begin by discussing the similarities between the numerical analyses (\autoref{sec:numerical_analyses}) and experimental validation (\autoref{sec:experimental_validation}). In all cases, the artificial target is successfully introduced to Bob. The effects of different combined strategies align well with the numerical analysis. The simplest combination, A1+B1, shows numerous target peaks. Both A1+B2 and A2+B2 produce similar RDMs, dominated by ridges. The third strategy, A2+B1, generates the cleanest RDM, showing mainly the artificial target. 

However, significant differences emerge between numerical and experimental results due to real-world environments and hardware limitations. First, static objects introduce clutter centered at zero-Doppler, along with ghost targets of moving entities. Since the clutter patterns differ between Alice-Bob and Eve-Bob, Bob may detect these discrepancies and react accordingly. Second, Eve’s range/Doppler-shifted signals for the artificial target cause clutter peaks to shift according to the artificial target’s parameters. Third, the experimental noise floor is higher than in simulations, which consider only thermal noise. Fourth, if the number of ridges--determined by mobile target peaks in the Alice-Bob channel--is high, the artificial target is more likely to be hidden by them. Lastly, when Eve uses A2 to form a beam towards Bob and null towards the target, calibration errors create imperfections in the radiation pattern, making mobile target peaks slightly visible.

\section{Conclusion}
\label{sec:conclusion}
This paper examines the design and application of target-spoofing deceptive jammers in WLAN sensing, highlighting vulnerabilities in communication-centric OFDM-based JCAS systems that rely on RDM processing. Through qualitative and quantitative analysis, we demonstrate how techniques like overcrowding, selective target injection, and forced synchronization disrupt target detection. Experiments confirm that low-cost SDR platforms can effectively carry out these jamming techniques without sophisticated hardware.

\review{We identify key vulnerabilities in RDM-based WLAN sensing: i) standardized OFDM symbols, ii) joint time and frequency synchronization, and iii) unprotected over-the-air negotiation. These are exploitable when i) JSR is above 8-12 dB, and ii) the CFO difference $\Delta$ is at least 3 ppm (or 6 ppm with wider subcarrier spacing). Under these conditions, target spoofing and deceptive jamming are highly effective.} These results underscore the need to address such weaknesses in future WLAN sensing and JCAS systems as they enter critical applications.

In summary, this study reveals substantial risks posed by target-spoofing and deceptive jamming, encouraging future work on securing physical-layer waveforms, enhancing synchronization, and developing adaptive methods to mitigate these vulnerabilities in OFDM-based JCAS systems.

\appendices

\section{Joint Time and Frequency Synchronization}
\label{app:jtfs}
Bob exploits a dominant LOS for JTFS, meaning ${\alpha_0\gg\alpha_p, \forall p}$ and ${\tau_0<\tau_p, \forall p}$.  The received LOS signal sampled at time $t=nT+mT_s$ (where $n$ and $m$ denote the fast and slow time indices, respectively) is expressed as \cite{vandebeek99,schmidl97}
\begin{align}
    r[n,m] = \alpha_0 s[n-n_0,m] e^{j2\pi\eta (nT+mT_s)}\nonumber
\end{align}
where the noise is omitted, $\eta$ represents the CFO between Alice and Bob, and $s[n,m]$ is the transmitted signal. For coarse synchronization, Bob computes the lag-1 auto-correlation over a sliding window of the received samples, expressed as
\begin{align}
    \Xi[k] = \sum_{n=-Q_{cp}}^{Q} r[n+k,m]r^*[n+k,m+1].\nonumber
\end{align}
Substituting the received signal into the equation yields the strongest peak at $k=n_0$ corresponding to the arrival time of the LOS path since ${\alpha_0\gg\alpha_p, \forall p}$. Assuming that the transmitted signal is normalized to ${\sum_{n=-Q_{cp}}^{Q}s[n,m]=1}$, the complex amplitude of this peak is defined as
\begin{align}
    \Xi[n_0] = \abs{\alpha_0}^2 e^{-j2\pi \eta T_s}.\nonumber
\end{align}
Here, the phase of $\Xi[n_0]$ is directly linked to the CFO between Alice and Bob. Hence, Bob can synchronize to the timing of the LOS based on the peak at ${k=n_0}$, and estimate and compensate the CFO using the phase of the peak ${\hat{\eta} = {\angle \Xi[n_0]}/{2\pi T_s}}$. This process ensures that Bob is correctly synchronized in both time and frequency domains.


\section{Round-trip-time for distance estimation}
\label{app:rtt}
\begin{figure}[h!]
    \centering
    \includegraphics[width=\linewidth]{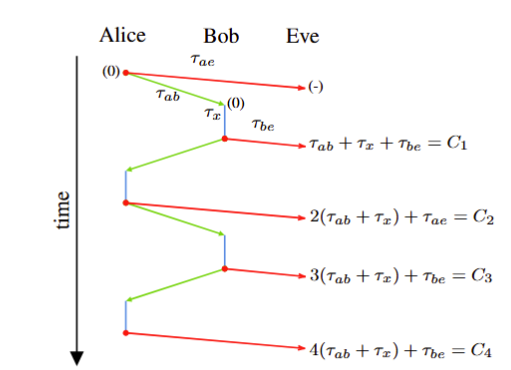}
    \caption{RTT-based distance estimation procedure. The green and red arrows correspond to the propagation delays over the related distances. Blue lines correspond to the standardized idle time $\tau_x$.}
    \label{fig:rtt_based_distance_estimation}
\end{figure}

The devices in the legitimate WLAN sensing network, Alice and Bob, estimate ${\tau_{ab}=d_{ab}/c}$ for themselves where $c$ is the speed of light. The process is visualized in \autoref{fig:rtt_based_distance_estimation} and summarized as follows \cite{mirkovic18}. I) Alice transmits and starts its reference clock. II) Bob receives the signal and starts its reference clock. III) Bob waits for the standardized duration of $\tau_x$, then transmits. IV) Alice receives the signal after ${2\tau_{ab}+\tau_x}$ seconds, then estimates $\tau_{ab}$. V) Alice waits $\tau_x$ seconds, then transmits. VI) Bob receives the signal after ${3\tau_{ab}+2\tau_x}$ seconds, and estimates $\tau_{ab}$. VII) Both devices now have $\tau_{ab}$.

Meanwhile, Eve can estimate ${\tau_{be}=d_{be}/c}$, ${\tau_{ae}=d_{ae}/c}$ and ${\tau_{ab}=d_{ab}/c}$ as follows \cite{jiang02}. I) Alice transmits. II) Eve receives the signal and knows that Alice has transmitted it. Eve does not start its reference clock yet and waits for the next transmission. III) Bob waits $\tau_x$ seconds, then transmits. IV) Eve receives Bob's signal and starts its reference clock since it knows that ${\tau_{ab}+\tau_x+\tau_{be}}$ seconds have passed. V) Eve listens for the next transmissions and keeps measurement times as $C_1$ to $C_4$. Once all the reception times are collected, the propagation delays are calculated as
\begin{align}
    \tau_{abx}=(C_3-C_1)/2,
    \tau_{be}=(3C_1-C_3)/2,
    \tau_{ae}=C_2\!\!-\!\!C_3\!+\!C_1 \nonumber
\end{align}
where $\tau_{abx}=\tau_{ab}+\tau_x$.

\balance 

\bibliographystyle{IEEEtran}
\bibliography{main}

\begin{thebibliography}{10}
\providecommand{\url}[1]{#1}
\csname url@samestyle\endcsname
\providecommand{\newblock}{\relax}
\providecommand{\bibinfo}[2]{#2}
\providecommand{\BIBentrySTDinterwordspacing}{\spaceskip=0pt\relax}
\providecommand{\BIBentryALTinterwordstretchfactor}{4}
\providecommand{\BIBentryALTinterwordspacing}{\spaceskip=\fontdimen2\font plus
\BIBentryALTinterwordstretchfactor\fontdimen3\font minus \fontdimen4\font\relax}
\providecommand{\BIBforeignlanguage}[2]{{%
\expandafter\ifx\csname l@#1\endcsname\relax
\typeout{** WARNING: IEEEtran.bst: No hyphenation pattern has been}%
\typeout{** loaded for the language `#1'. Using the pattern for}%
\typeout{** the default language instead.}%
\else
\language=\csname l@#1\endcsname
\fi
#2}}
\providecommand{\BIBdecl}{\relax}
\BIBdecl

\bibitem{du21}
R.~Du, H.~Hua, H.~Xie, X.~Song, Z.~Lyu, M.~Hu, Y.~Xin, S.~McCann, M.~Montemurro, T.~X. Han \emph{et~al.}, ``An overview on {IEEE 802.11 bf: WLAN} sensing,'' \emph{arXiv preprint arXiv:2207.04859}, 2021.

\bibitem{zhang21a}
J.~A. Zhang, M.~L. Rahman, K.~Wu, X.~Huang, Y.~J. Guo, S.~Chen, and J.~Yuan, ``Enabling joint communication and radar sensing in mobile networks—a survey,'' \emph{IEEE Communications Surveys \& Tutorials}, vol.~24, no.~1, pp. 306--345, 2021.

\bibitem{wlan_sensing_scenarios}
{IEEE 802.11bf TG}, ``{IEEE WLAN} sensing use cases, official document,'' \url{https://mentor.ieee.org/802.11/dcn/20/11-20-1712-02-00bf-wifi-sensing-use-cases.xlsx}, last accessed: 17/10/2023.

\bibitem{zhang21b}
Z.~Zhang and M.~Krunz, ``Preamble injection and spoofing attacks in {Wi-Fi} networks,'' in \emph{2021 IEEE Global Communications Conference (GLOBECOM)}.\hskip 1em plus 0.5em minus 0.4em\relax IEEE, 2021, pp. 1--6.

\bibitem{lapan13}
M.~J. La~Pan, T.~C. Clancy, and R.~W. McGwier, ``Phase warping and differential scrambling attacks against ofdm frequency synchronization,'' in \emph{2013 IEEE International Conference on Acoustics, Speech and Signal Processing}, 2013, pp. 2886--2890.

\bibitem{lapan12}
M.~J.~L. Pan, T.~C. Clancy, and R.~W. McGwier, ``Jamming attacks against ofdm timing synchronization and signal acquisition,'' in \emph{MILCOM 2012 - 2012 IEEE Military Communications Conference}, 2012, pp. 1--7.

\bibitem{clancy11}
T.~C. Clancy, ``Efficient ofdm denial: Pilot jamming and pilot nulling,'' in \emph{2011 IEEE International Conference on Communications (ICC)}, 2011, pp. 1--5.

\bibitem{patwardhan14}
G.~Patwardhan and D.~Thuente, ``Jamming beamforming: A new attack vector in jamming {IEEE 802.11} ac networks,'' \emph{2014 IEEE Military Communications Conference}, pp. 1534--1541, 2014.

\bibitem{zhao19}
S.~Zhao, Z.~Lu, Z.~Luo, and Y.~Liu, ``Orthogonality-sabotaging attacks against ofdma-based wireless networks,'' in \emph{IEEE INFOCOM 2019 - IEEE Conference on Computer Communications}, 2019, pp. 1603--1611.

\bibitem{merwe18}
J.~R. v.~d. Merwe, X.~Zubizarreta, I.~Lukčin, A.~Rügamer, and W.~Felber, ``Classification of spoofing attack types,'' in \emph{2018 European Navigation Conference (ENC)}.\hskip 1em plus 0.5em minus 0.4em\relax IEEE, 2018, pp. 91--99.

\bibitem{matte15}
C.~Matte, J.~P. Achara, and M.~Cunche, ``Device-to-identity linking attack using targeted {Wi-Fi} geolocation spoofing,'' \emph{Proceedings of the 8th ACM Conference on Security \& Privacy in Wireless and Mobile Networks}, pp. 1--6, 2015.

\bibitem{tippenhauer09}
N.~O. Tippenhauer, K.~B. Rasmussen, C.~P{\"o}pper, and S.~{\v{C}}apkun, ``Attacks on public {WLAN}-based positioning systems,'' \emph{Proceedings of the 7th international conference on Mobile systems, applications, and services}, pp. 29--40, 2009.

\bibitem{liu17}
D.~Liu, Y.~Xu, and X.~Huang, ``Identification of location spoofing in wireless sensor networks in non-line-of-sight conditions,'' \emph{IEEE Transactions on Industrial Informatics}, vol.~14, no.~6, pp. 2375--2384, 2017.

\bibitem{pirayesh22}
H.~Pirayesh and H.~Zeng, ``Jamming attacks and anti-jamming strategies in wireless networks: A comprehensive survey,'' \emph{IEEE communications surveys \& tutorials}, vol.~24, no.~2, pp. 767--809, 2022.

\bibitem{gunther14}
C.~G{\"u}nther, ``A survey of spoofing and counter-measures,'' \emph{NAVIGATION: Journal of the Institute of Navigation}, vol.~61, no.~3, pp. 159--177, 2014.

\bibitem{melki19}
R.~Melki, H.~N. Noura, M.~M. Mansour, and A.~Chehab, ``A survey on {OFDM} physical layer security,'' \emph{Physical Communication}, vol.~32, pp. 1--30, 2019.

\bibitem{schuerger08}
J.~Schuerger and D.~Garmatyuk, ``Deception jamming modeling in radar sensor networks,'' \emph{IEEE Military Communications Conference}, pp. 1--7, 2008.

\bibitem{schuerger09}
------, ``Performance of random {OFDM} radar signals in deception jamming scenarios,'' \emph{IEEE Radar Conference}, pp. 1--6, 2009.

\bibitem{tan21}
M.~Tan, C.~Wang, B.~Xue, and J.~Xu, ``A novel deceptive jamming approach against frequency diverse array radar,'' \emph{IEEE Sensors Journal}, vol.~21, no.~6, pp. 8323--8332, 2021.

\bibitem{sun18}
Q.~Sun, T.~Shu, K.-B. Yu, and W.~Yu, ``Efficient deceptive jamming method of static and moving targets against sar,'' \emph{IEEE Sensors Journal}, vol.~18, no.~9, pp. 3610--3618, 2018.

\bibitem{ji24}
P.~Ji, S.~Xing, D.~Dai, B.~Pang, and D.~Feng, ``A smart multitransmitter cooperative false images generation method against multichannel sar-gmti,'' \emph{IEEE Transactions on Geoscience and Remote Sensing}, vol.~62, pp. 1--17, 2024.

\bibitem{yang22}
K.~Yang, F.~Ma, D.~Ran, W.~Ye, and G.~Li, ``Fast generation of deceptive jamming signal against spaceborne sar based on spatial frequency domain interpolation,'' \emph{IEEE Transactions on Geoscience and Remote Sensing}, vol.~60, pp. 1--15, 2022.

\bibitem{drfm_main}
S.~Roome, ``Digital radio frequency memory,'' \emph{Electronics \& communication engineering journal}, 1990.

\bibitem{liang20}
Y.~Liang, J.~Ren, and T.~Li, ``Secure {OFDM} system design and capacity analysis under disguised jamming,'' \emph{IEEE Transactions on Information Forensics and Security}, vol.~15, pp. 738--752, 2020.

\bibitem{fang24}
X.~Fang, M.~Li, S.~Li, D.~Ramaccia, A.~Toscano, F.~Bilotti, and D.~Ding, ``Diverse frequency time modulation for passive false target spoofing: Design and experiment,'' \emph{IEEE Transactions on Microwave Theory and Techniques}, vol.~72, no.~3, pp. 1932--1942, 2024.

\bibitem{srinivasan24}
M.~Srinivasan, L.~Senigagliesi, H.~Chen, A.~Chorti, M.~Baldi, and H.~Wymeersch, ``{AoA}-based physical layer authentication in analog arrays under impersonation attacks,'' \emph{arXiv preprint arXiv:2407.08282}, 2024.

\bibitem{varotto24a}
M.~Varotto, S.~Valentin, and S.~Tomasin, ``Detecting {5G} signal jammers using spectrograms with supervised and unsupervised learning,'' \emph{arXiv preprint arXiv:2405.10331}, 2024.

\bibitem{varotto23}
------, ``Detecting {5G} signal jammers with autoencoders based on loose observations,'' in \emph{2023 IEEE Globecom Workshops (GC Wkshps)}.\hskip 1em plus 0.5em minus 0.4em\relax IEEE, 2023, pp. 160--165.

\bibitem{varotto24b}
M.~Varotto, F.~Heinrichs, T.~Schuerg, S.~Tomasin, and S.~Valentin, ``Detecting {5G} narrowband jammers with {CNN}, k-nearest neighbors, and support vector machines,'' \emph{arXiv preprint arXiv:2405.09564}, 2024.

\bibitem{li24a}
J.~Li and U.~Mitra, ``Channel state information-free location-privacy enhancement: Delay-angle information spoofing,'' in \emph{ICC 2024 - IEEE International Conference on Communications}, 2024, pp. 3767--3772.

\bibitem{li24b}
------, ``Channel state information-free location-privacy enhancement: Fake path injection,'' \emph{IEEE Transactions on Signal Processing}, vol.~72, pp. 3745--3760, 2024.

\bibitem{argyriou23}
A.~Argyriou, ``Range-doppler spoofing in {OFDM} signals for preventing wireless passive emitter tracking,'' in \emph{2023 IEEE Radar Conference (RadarConf23)}.\hskip 1em plus 0.5em minus 0.4em\relax IEEE, 2023, pp. 1--6.

\bibitem{yildirim24}
H.~C. Yildirim, M.~F. Keskin, H.~Wymeersch, and F.~Horlin, ``Deceptive jamming in {WLAN} sensing,'' \emph{2024 IEEE Radar Conference (RadarConf24)}, 2024.

\bibitem{ropitault23}
T.~Ropitault, C.~da~Silva, S.~Blandino, A.~Sahoo, N.~Golmie, K.~Yoon, C.~Aldana, and C.~Hu, ``{IEEE 802.11 bf WLAN} sensing procedure: Enabling the widespread adoption of {Wi-Fi} sensing,'' \emph{IEEE Communications Standards Magazine}, 2023.

\bibitem{bejarano13}
O.~Bejarano, E.~W. Knightly, and M.~Park, ``{IEEE} 802.11 ac: from channelization to multi-user {MIMO},'' \emph{IEEE Communications Magazine}, 2013.

\bibitem{vandebeek99}
J.-J. van~de Beek, P.~Borjesson, M.-L. Boucheret, D.~Landstrom, J.~Arenas, P.~Odling, C.~Ostberg, M.~Wahlqvist, and S.~Wilson, ``A time and frequency synchronization scheme for multiuser {OFDM},'' \emph{IEEE Journal on Selected Areas in Communications}, vol.~17, no.~11, pp. 1900--1914, 1999.

\bibitem{schmidl97}
T.~Schmidl and D.~Cox, ``Robust frequency and timing synchronization for ofdm,'' \emph{IEEE Transactions on Communications}, vol.~45, no.~12, pp. 1613--1621, 1997.

\bibitem{horlin08}
F.~Horlin and A.~Bourdoux, \emph{Digital compensation for analog front-ends: a new approach to wireless transceiver design}.\hskip 1em plus 0.5em minus 0.4em\relax John Wiley \& Sons, 2008.

\bibitem{durgin}
G.~D. Durgin, \emph{Space-time wireless channels}.\hskip 1em plus 0.5em minus 0.4em\relax Prentice Hall Professional, 2003.

\bibitem{wei24}
Z.~Wei, J.~Jia, Y.~Niu, L.~Wang, H.~Wu, H.~Yang, and Z.~Feng, ``Integrated sensing and communication channel modeling: A survey,'' \emph{IEEE Internet of Things Journal}, pp. 1--1, 2024.

\bibitem{richards10}
M.~A. Richards, J.~A. Scheer, and W.~A. Holm, \emph{Principles of Modern Radar: Basic principles}.\hskip 1em plus 0.5em minus 0.4em\relax The Institution of Engineering and Technology, 2010.

\bibitem{chiueh12}
T.-D. Chiueh, P.-Y. Tsai, and I.-W. Lai, \emph{Baseband receiver design for wireless {MIMO-OFDM} communications}.\hskip 1em plus 0.5em minus 0.4em\relax John Wiley \& Sons, 2012.

\bibitem{mahmood16}
A.~Mahmood, R.~Exel, H.~Trsek, and T.~Sauter, ``Clock synchronization over {IEEE 802.11}—a survey of methodologies and protocols,'' \emph{IEEE Transactions on Industrial Informatics}, vol.~13, no.~2, pp. 907--922, 2016.

\bibitem{yildirim20}
H.~C. Yildirim, J.~Louveaux, P.~De~Doncker, and F.~Horlin, ``Impact of interference on {OFDM} based radars,'' \emph{IEEE Vehicular Technology Conference}, 2020.

\bibitem{MUSIC}
R.~Schmidt, ``Multiple emitter location and signal parameter estimation,'' \emph{IEEE Transactions on Antennas and Propagation}, vol.~34, no.~3, pp. 276--280, 1986.

\bibitem{ESPRIT}
R.~Roy and T.~Kailath, ``{ESPRIT}-estimation of signal parameters via rotational invariance techniques,'' \emph{IEEE Transactions on Acoustics, Speech, and Signal Processing}, vol.~37, no.~7, pp. 984--995, 1989.

\bibitem{boulic90}
R.~Boulic, N.~M. Thalmann, and D.~Thalmann, ``A global human walking model with real-time kinematic personification,'' \emph{The visual computer}, 1990.

\bibitem{blake88}
S.~Blake, ``{OS-CFAR} theory for multiple targets and nonuniform clutter,'' \emph{IEEE Transactions on Aerospace and Electronic Systems}, vol.~24, no.~6, pp. 785--790, 1988.

\bibitem{mirkovic18}
D.~Mirkovic, G.~Armitage, and P.~Branch, ``A survey of round trip time prediction systems,'' \emph{IEEE Communications Surveys \& Tutorials}, vol.~20, no.~3, pp. 1758--1776, 2018.

\bibitem{jiang02}
H.~Jiang and C.~Dovrolis, ``Passive estimation of {TCP} round-trip times,'' \emph{ACM SIGCOMM Computer Communication Review}, vol.~32, no.~3, pp. 75--88, 2002.

\end{thebibliography}


\end{document}